\documentclass[review,11pt]{elsarticle}
\usepackage{lineno,hyperref}
\modulolinenumbers[5]
\journal{Combustion and Flame}
\usepackage{epstopdf}
\usepackage{setspace}
\usepackage{graphicx}
\usepackage{color}
\usepackage{rotating}
\usepackage{soul}
\usepackage{geometry}

\biboptions{sort&compress}
\usepackage{caption}
\usepackage[table]{xcolor}
\usepackage{multirow}
\usepackage{arydshln}
\usepackage{amsmath,amssymb}
\usepackage{tikz}
\usetikzlibrary{calc,topaths}

\usepackage{chemfig}

\begin{document}

\begin{frontmatter}

\title{Graphene oxide doped ethanol droplet combustion: Ignition delay and contribution of atomization to burning rate}
\author{Sepehr Mosadegh$^{\mathrm{a}}$, Ahmadreza Ghaffarkhah$^{\mathrm{a}}$, Colin van der Kuur$^{\mathrm{b}}$, Mohammad Arjmand$^{\mathrm{a}}$, Sina Kheirkhah$^{\mathrm{a,*}}$}
\address{$^{\mathrm{a}}$School of Engineering, The University of British Columbia, Kelowna, British Columbia, Canada, V1V 1V7}
\address{$^{\mathrm{b}}$ZEN Graphene Solutions, 210-1205 Amber Dr., Thunder Bay, Ontario, Canada, P7B 6M4}
\cortext[mycorrespondingauthor]{Corresponding author (sina.kheirkhah@ubc.ca)}

\begin{abstract}

Effects of graphene oxide nanomaterials addition and oxidation level on ignition delay and burning rate of ethanol droplets are experimentally investigated. Three graphene oxide samples are synthesized and characterized. Separate high-speed $\mathrm{OH^*}$ chemiluminescence and high-speed shadowgraphy images are collected. The results suggest that increasing the loading concentration from 0 to 0.1\% (by weight) generally increases the ignition delay, except for ethanol doped with the highly oxidized graphene. The results show that unlike pure ethanol droplets, atomization may occur for the doped ethanol droplets. It is demonstrated that, independent of the tested conditions, atomization occurs in the second half of the droplet lifetime. The probability density function of the atomized baby droplet diameter, initial projected velocity, and length of the projected trajectory are similar for all tested conditions and independent of the oxidation level and loading concentration of the additives. The joint probability density function calculated for the atomization-related parameters against one another suggests that the majority of the baby droplets feature a relatively short lifetime, indicating they may potentially burn inside the flame envelope. Using droplet surface regression curves versus time, the burning rate for periods in which the atomization does not occur, and for the periods that the atomization is present are estimated. The former burning rate is shown to enhance by increasing the loading concentration and reducing the oxidation level of graphene. However, supported by Fourier-Transform Infrared spectroscopy of the graphene oxide samples, it is found that a maximum increase in the latter burning rate for both loading concentrations occurs for ethanol doped with the graphene oxide that features maximum amount of infrared radiation absorption. To quantify the effect of atomization on the droplet mass loss, a conservation of mass framework is utilized, and it is shown that relatively intense atomization suppresses the mass loss. Doping ethanol with graphene oxide and increasing the loading concentration from 0.01 to 0.1\% enhances the overall burning rate, with a maximum enhancement of 8.4\% pertaining to addition of reduced oxidized graphene oxide and for the loading concentration of 0.1\%.

\end{abstract}


\end{frontmatter}

\section{Introduction}
\label{Introduction}

Combustion characteristics of droplets doped with both carbonaceous\cite{gan2012optical,ghamari2017combustion,aboalhamayie2019evaporation,khond2017experimental,tanvir2017evaporation,ooi2017improving,ooi2016graphite,yadav2019combustion,singh2019effect,singh2020effect,tanvir2016droplet,sabourin2009functionalized,bennewitz2020combustion,murr2004carbon,miyasaka1980combustion,szekely1982combustion,yao1983behavior,szekely1984effects,sakai1985single,clausen1988effects,lee1991gasification,wong1992microexplosion,bhadra1997ignition} and non-carbon-based \cite{yetter2009metal,sim2018effects,javed2014effects,dai2019experimental,javed2013evaporation,guerieri2020droplet,javed2015autoignition,bello2018surface,tanvir2015effect,bello2015reaction,gan2011combustion,gan2012combustion,miglani2014insight,miglani2015coupled,pandey2019high,miglani2014suppression,miglani2015effect,taghavifar2020application,heidari2020novel} additives have been investigated in the past. Results of past studies show the addition of nanomaterials to liquid fuels can potentially allow for improving some thermofluidic characteristics, such as thermal conductivity~\cite{gan2012optical,ghamari2017combustion,aboalhamayie2019evaporation,jang2004role,prasher2005thermal,yu2003role,liang2011thermal,eastman2004thermal,timofeeva2007thermal,wu2010effect,taylor2013small,keblinski2005nanofluids,mahian2019recent,tawfik2017experimental}, which can lead to enhanced evaporation rate \cite{aboalhamayie2019evaporation,khond2017experimental,tanvir2017evaporation,javed2014effects,dai2019experimental,javed2013evaporation} and burning rate  \cite{gan2012optical,ghamari2017combustion,ooi2017improving,ooi2016graphite,yadav2019combustion,singh2019effect,singh2020effect,tanvir2016droplet,sabourin2009functionalized,sim2018effects,guerieri2020droplet,javed2015autoignition,bello2018surface,tanvir2015effect,bello2015reaction}. Past investigations~\cite{javed2014effects,gan2011combustion,gan2012combustion,miglani2014insight,miglani2015coupled,pandey2019high,basu2016combustion,yao2021atomization} show that complex thermofluidic phenomena, specifically atomization, can significantly influence droplet combustion. Although several studies reported the occurrence of atomization, see for example \cite{ghamari2017combustion,ooi2017improving,singh2019effect,singh2020effect}, our quantitative understanding related to the characteristics of this phenomenon is limited and requires further investigations. For example, trajectory and velocity of the ejected baby droplets as a result of atomization, as well as how such phenomenon influences the droplet burning rate is of interest. In the following, background related to the effects of nanomaterials addition on both liquid fuel thermophysical properties and combustion characteristics are briefly reviewed. \par

Addition of nanomaterials to liquid fuels can change conductive, convective, and radiative heat transfer to the droplet, which, in turn, influences the droplet evaporation and burning rate \cite{gan2012optical,ghamari2017combustion,aboalhamayie2019evaporation,tanvir2017evaporation,ooi2017improving,ooi2016graphite,singh2020effect,tanvir2016droplet,sabourin2009functionalized,basu2016combustion}. Thermal conductivity is highly influenced by the type of the added particles and their loading concentration, temperature of the fuel and particles suspension, and the added particle size. Generally, increasing the loading concentration \cite{taylor2013small,keblinski2005nanofluids,mahian2019recent}, increasing the suspension temperature \cite{mahian2019recent,tawfik2017experimental}, and decreasing the size of the nanomaterial particles \cite{mahian2019recent,tawfik2017experimental} are expected to increase the thermal conductivity. For example, Agarwal et al. \cite{agarwal2016experimental} and Aboalhamayie et al. \cite{aboalhamayie2019evaporation} reported that adding 0.2 and 3\% by weight of carbon nanotubes and graphene nanoplatelets to jet fuel and kerosene may increase the thermal conductivity by 23\% and 29\%, respectively. The increase of the thermal conductivity by addition of nanomaterials is due to enhancement of the Brownian motion \cite{jang2004role,prasher2005thermal} and generation of solid-liquid interface thermal bridge \cite{yu2003role,liang2011thermal,eastman2004thermal} inside the droplet. However, excessive doping of the liquid fuels with nanomaterials may lead to formation of aggregates, which may or may not increase the fuel thermal conductivity. For example, studies of~\cite{liang2011thermal,timofeeva2007thermal,wu2010effect} reported that nanomaterials aggregation increases the thermal conductivity, while studies of \cite{ghamari2017combustion,tawfik2017experimental} suggest nanomaterials aggregation decreases the thermal conductivity. Compared to the conductive heat transfer, the convection heat transfer inside the doped liquid fuel droplets is improved as a result of circulatory flow inside the droplets which influences/is influenced by aggregation of the nanomaterials and their morphology~\cite{ghamari2016experimental}. Radiation heat transfer can also be improved by the addition of nanomaterials \cite{gan2012optical,ghamari2017combustion,tanvir2017evaporation,ooi2017improving,ooi2016graphite,singh2020effect,tanvir2016droplet}. Majority of liquid fuels are transparent to the radiation wavelengths emitted from the flame. However, the addition of carbonaceous nanomaterials decreases the droplet optical transmittance, which increases the radiation absorption~ \cite{gan2012optical,ghamari2017combustion,tanvir2016droplet}. The amount of increase depends on the emission wavelength and the droplet size, as shown in, for example \cite{tanvir2016droplet}. \par

Two combustion characteristics, specifically, ignition delay and burning rate, are of relevance to the present study. Past investigations \cite{ooi2017improving,ooi2016graphite,singh2019effect,bennewitz2020combustion} show some controversy in the reported results related to the effect of nanomaterials addition on liquid fuel ignition delay. On one hand, Ooi et al. \cite{ooi2017improving,ooi2016graphite} reported that the ignition delay of diesel and biodiesel decreases 46.5\% by addition of 0.01--0.1\% by weight graphite oxide. On the other hand, Bennewitz et al. \cite{bennewitz2020combustion} reported adding 1\% by weight of graphene nanomaterials increases the ignition delay of rocket-grade kerosene by 20\%. Similar to Benewitz et al. \cite{bennewitz2020combustion}, Singh et al. \cite{singh2019effect} reported that ignition delay increases about 3\% and 10\% by adding 3\% by weight of acetylene black nanomaterials to petrodiesel and biodiesel, respectively. The reason for such controversy may be related to the different combustion chemistry of the studied fuel, and is not the focus of the present investigation. \par

Addition of nanomaterials to liquid fuels alters the heat transfer to/from the droplet, which influences the droplet evaporation rate, see for example \cite{aboalhamayie2019evaporation,khond2017experimental,tanvir2017evaporation,javed2014effects,dai2019experimental,javed2013evaporation}. Results of Tanvir et al.~\cite{tanvir2017evaporation} suggest that the circulatory flow inside the droplet accumulates the nanomaterials at the droplet surface, creating a porous shell. This shell formation may either positively or negatively impact the evaporation rate. On one hand, the shell formation decreases the convection heat transfer from the surrounding to the liquid, which limits the enhancement of the evaporation rate by the addition of the nanomaterials~\cite{basu2016combustion}. Also, once the porous shell is formed at the droplet surface, the liquid fuel permeates from the droplet core through the porous shell and to the droplet surface. This phenomenon may negatively influence the evaporation rate. On the other hand, the addition of the nanomaterials can lead to the droplet temperature rise~\cite{basu2016combustion}, which enhances permeation of the liquid fuel through the porous shell~\cite{pandey2019high} and potentially the evaporation rate. Both the formation of the porous shell as well as the droplet temperature rise may lead to competing effects on the evaporation rate, and the resultant effect depends on the type of the utilized nanomaterials and the amount of the nanomaterials doping. For example, Gan and Qiao \cite{gan2012optical} used a mercury lamp and investigated the effect of radiation absorption on the evaporation rate. Their results suggest that, for a fixed loading concentration of 0.1\% by weight and at a radiation intensity of 175~W, the addition of multi-walled carbon nanotubes to ethanol increased the evaporation rate by 25.8\%. However, this increase was 5.7\% for ethanol doped with carbon nanoparticles for an identical lamp radiation intensity. \par 

The effect of nanomaterials addition on the fuel burning rate has been investigated in several past studies, see for example~\cite{gan2012optical,ghamari2017combustion,ooi2017improving,ooi2016graphite,yadav2019combustion,singh2019effect,singh2020effect,tanvir2016droplet,sabourin2009functionalized,sim2018effects,guerieri2020droplet,javed2015autoignition,bello2018surface,tanvir2015effect,bello2015reaction}. Results of Ghamari et al.~\cite{ghamari2017combustion} suggest that there exists an optimum loading concentration at which the burning rate is maximized. Specifically, up to about 10\% enhancement of jet fuel burning rate was reported by adding 1.5, 0.25, and 0.1\% by weight of activated carbon nanoparticles, multi-walled carbon nanotubes, and graphene nanomaterials, respectively. Similarly, Ooi et al. \cite{ooi2017improving,ooi2016graphite}, Singh et al.~\cite{singh2019effect}, Yadav et al. \cite{yadav2019combustion}, and Sabourin et al. \cite{sabourin2009functionalized} added 0.01--0.1\% graphite oxide, 2\% acetylene black nanomaterials, 0.2\% graphene nanomaterials, and 0.5\% by weight functionalized graphene sheets into diesel, biodisel, rocket-grade kerosene, and nitromethane; and, they reported about 29, 13, 14, and 175\% increase in the burning rate, respectively. Decreasing the nanomaterial size can increase the burning rate of the liquid fuels. For example, Tanvir et al.~\cite{tanvir2016droplet} showed that decreasing the graphite nanomaterials size from 100 to 50~nm can increase the burning rate of ethanol doped droplets by about 64\%. Compared to the effect of decreasing the nanomaterials size on the burning rate, excessive doping~\cite{ghamari2017combustion} and poor suspension stability may lead to a decrease of the burning rate. \par

Two mechanisms can influence the burning rate of the nanomaterials doped droplets~\cite{gan2012optical,ghamari2017combustion,aboalhamayie2019evaporation,tanvir2017evaporation,ooi2017improving,ooi2016graphite,singh2020effect,tanvir2016droplet,sabourin2009functionalized,gan2011combustion,gan2012combustion,miglani2014insight,miglani2015coupled,pandey2019high,miglani2015effect,basu2016combustion,wang2019experimental,law2010combustion}. First, doping with nanomaterials influences the fuel thermal conductivity, radiation absorption, as well as the permeation of liquid fuel through the nanomaterial shell formed at the droplet surface. Such influences can potentially increase the evaporation rate and as a result the droplet burning rate~\cite{gan2012optical,ghamari2017combustion,aboalhamayie2019evaporation,tanvir2017evaporation,ooi2017improving,ooi2016graphite,singh2020effect,tanvir2016droplet,sabourin2009functionalized,basu2016combustion}. Second, existence of nanomaterials leads to formation of bubbles inside the droplet, as a result of nucleation occurring at nanomaterial-liquid interface \cite{ghamari2017combustion,javed2014effects,gan2012combustion,miglani2014insight,basu2016combustion,wang2019experimental}. After the bubbles are formed, the gas pressure inside the bubbles rapidly increases due to surface regression of the main droplet, heat transfer from the flame, and/or bubbles coalescence~\cite{basu2016combustion}. Then, the bubbles reach the droplet surface, generate surface corrugations, and rupture the surface. This is followed by ligaments formation and its break-up into smaller baby droplets, which may burn later in the flame \cite{gan2011combustion,gan2012combustion,miglani2014insight,miglani2015coupled,pandey2019high,miglani2015effect,law2010combustion}. This process (which is referred to as micro-explosion) may take place once. However, as soon as the baby droplets are ejected, corrugations may be generated on the main droplet surface, increasing the surface area, enhancing the heat transfer to the droplet, producing new bubbles, and, finally, ejecting new baby droplets. This process can repeat several times during the main droplet lifetime and is referred to as the atomization \cite{miglani2015coupled}. \par

Although several investigations have reported formation and ejection of the baby droplets during doped droplet combustion, several questions remain to be addressed. First, effect of the graphene oxide addition on the ethanol droplets ignition delay is yet to be studied. Second, effects of baby droplets ejections on the main droplet evaporation and overall droplet burning rate are not quantified to the best knowledge of the authors. The present study follows three objectives. First, we aim to measure the ignition delay of the ethanol droplets doped with the graphene oxide. The second objective of the present investigations is to characterize the baby droplets size, trajectory of motion, initial velocity of ejection, and length of travel, as well as how/if these are influenced by the addition of graphene oxide nanomaterials. The third objective of the present investigation is to study the effects of such baby droplets formation and ejection on the evaporation mass loss, the overall droplet burning rate, and how these are influenced by the addition of the graphene oxide nanomaterials. In the following, a detailed characterization of the doping agent is provided in section~\ref{Nanofuel}. Then, the experimental methodology, results, and conclusions are presented in sections~\ref{Methodology},~\ref{results}, and~\ref{conclusion}, respectively. \par

\section{Preparation and characterization of the doped liquid fuel}
\label{Nanofuel}

Graphene oxide (GO) is used as the doping agent in the present investigation. While allowing to improve several liquid fuel thermophysical properties, laboratory-grade graphene oxide is relatively inexpensive to synthesize~\cite{sabourin2009functionalized,sabourin2010exploring,chen2020effect}. GO properties and how they influence the droplet combustion characteristics are strongly dependent on the existing functional groups~\cite{chen2020effect,mcallister2007single,sim2015understanding,schniepp2006functionalized,sabourin2009functionalized,ooi2017improving,javed2015autoignition}. Thus, extensive characterization of GO is important and is conducted in the present investigation. Here, three types of GO were synthesized and suspended in ethanol. A high purity (more than 99\%) graphite powder, referred to as Albany graphite \cite{zeng2019effects}, was used to synthesize highly oxidized graphene oxide (hGO). Then, hGO was thermally treated using a vacuum-assisted oven at 120~$\mathrm{^oC}$ with separate short and long treatment time periods, generating partially oxidized graphene oxide (pGO) and reduced oxidized graphene oxide (rGO), respectively. Details related to synthesis and characterization of these additives are discussed in the following. \par

First, 10~g of graphite powder was soaked inside 300~mL of $\mathrm{H_{2}SO_{4}}$ (with 98\% purity), and the suspension was stirred at 50~$\mathrm{^oC}$ with a rotation speed of 300~rpm for 18 hours. Then, the suspension container was positioned inside an ice bath, reducing its temperature to 5~$\mathrm{^oC}$. Next, 40~g of $\mathrm{KMnO_{4}}$ was gradually (about 0.7~g/min) added to the suspension. Then, the suspension was stirred, increasing its temperature to 50~$\mathrm{^oC}$. At this temperature, the suspension was stirred for 12 hours. Then, the suspension was positioned inside an ice bath, and 600~mL of distilled water (at room temperature) was added to the suspension. After this, diluted hydrogen peroxide (30\% $\mathrm{H_{2}O_{2}}$ and 70\% by weight distilled water) was added slowly (one droplet at a time) to the suspension until its colour changed from dark brown to yellow. Finally, in order to remove impurities, e.g., $\mathrm{SO_{4}^{2+}}$ and $\mathrm{K^{+}}$, the suspension was washed two times with a 1~M $\mathrm{HCl}$ solution (1 mole of $\mathrm{HCl}$ in 1 liter of $\mathrm{HCl}$ and water), two times with distilled water, and two times with ethanol, using the Eppendorf-5804 centrifuge operating at a rotation speed of 11000~rpm. The final sediments are dried using a temperature-controlled surface (SCILOGEX) at 60~$\mathrm{^oC}$ for 12 hours. As will be discussed later in this subsection, the product features the maximum oxygen content and is referred to as hGO. \par

\begin{figure}[!t]
	\centering
	\includegraphics[width=0.5\textwidth]{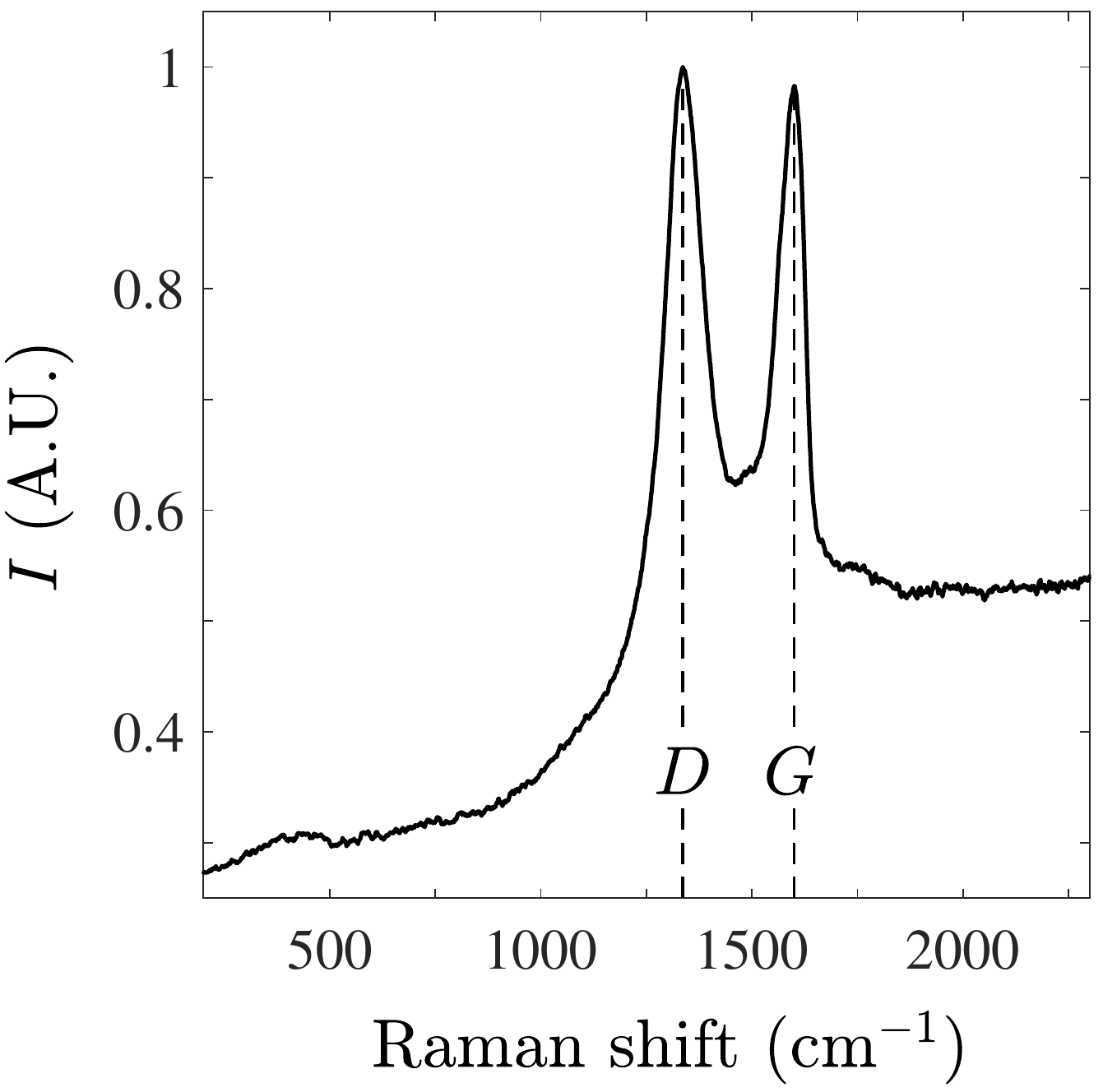} 
	\caption{Raman spectrum of the synthesized hGO.}
	\label{fig:Raman}
\end{figure}

Raman spectroscopy, similar to \cite{haddadi2021zinc}, was performed using the First guard Raman spectrometer (from Rigaku) to assess characteristics of the synthesized graphene oxide. The Raman shift was varied from 100 to 2300 $\mathrm{cm^{-1}}$ with a resolution of 0.65 $\mathrm{cm^{-1}}$. Figure~\ref{fig:Raman} shows the variation of the Raman scattering intensity normalized by the corresponding maximum versus the Raman shift wave-number. Two peaks, highlighted by $D$ and $G$ are detected, which correspond to the wave-numbers of 1336 and 1601 $\mathrm{cm^{-1}}$, respectively. The peak occurring at $D$ pertains to the defects in the structure of GO, likely due to the presence of hydroxyl and epoxide groups \cite{akhavan2015bacteriorhodopsin,haddadi2018fabrication}. The peak highlighted by $G$ pertains to the planar vibration mode of the carbon atoms~\cite{haddadi2018fabrication}. The Raman spectrum is similar to that previously reported in past investigations related to graphene oxide, see for example \cite{gopinath2018characterization,marcano2010improved,haddadi2018fabrication,nourbakhsh2010bandgap} and confirm the synthesized hGO material indeed features the molecular structure of graphene oxide. \par

\begin{figure}[!t]
	\centering
	\includegraphics[width=1\textwidth]{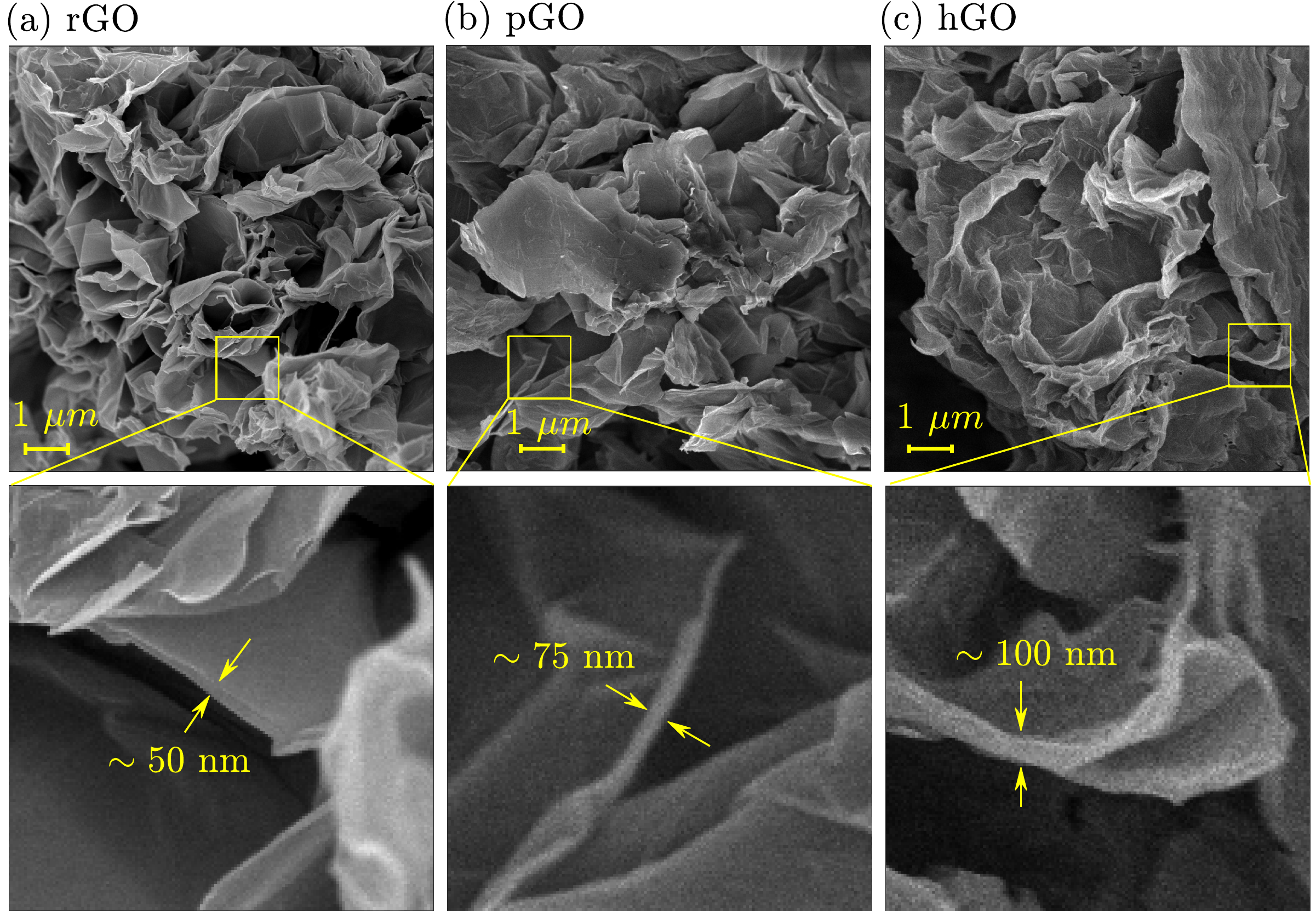} 
	\caption{High-resolution SEM images of (a) rGO, (b) pGO, and (c) hGO, respectively.}
	\label{fig:SEM}
\end{figure}

\begin{figure}[!t]
	\centering
	\includegraphics[width=0.5\textwidth]{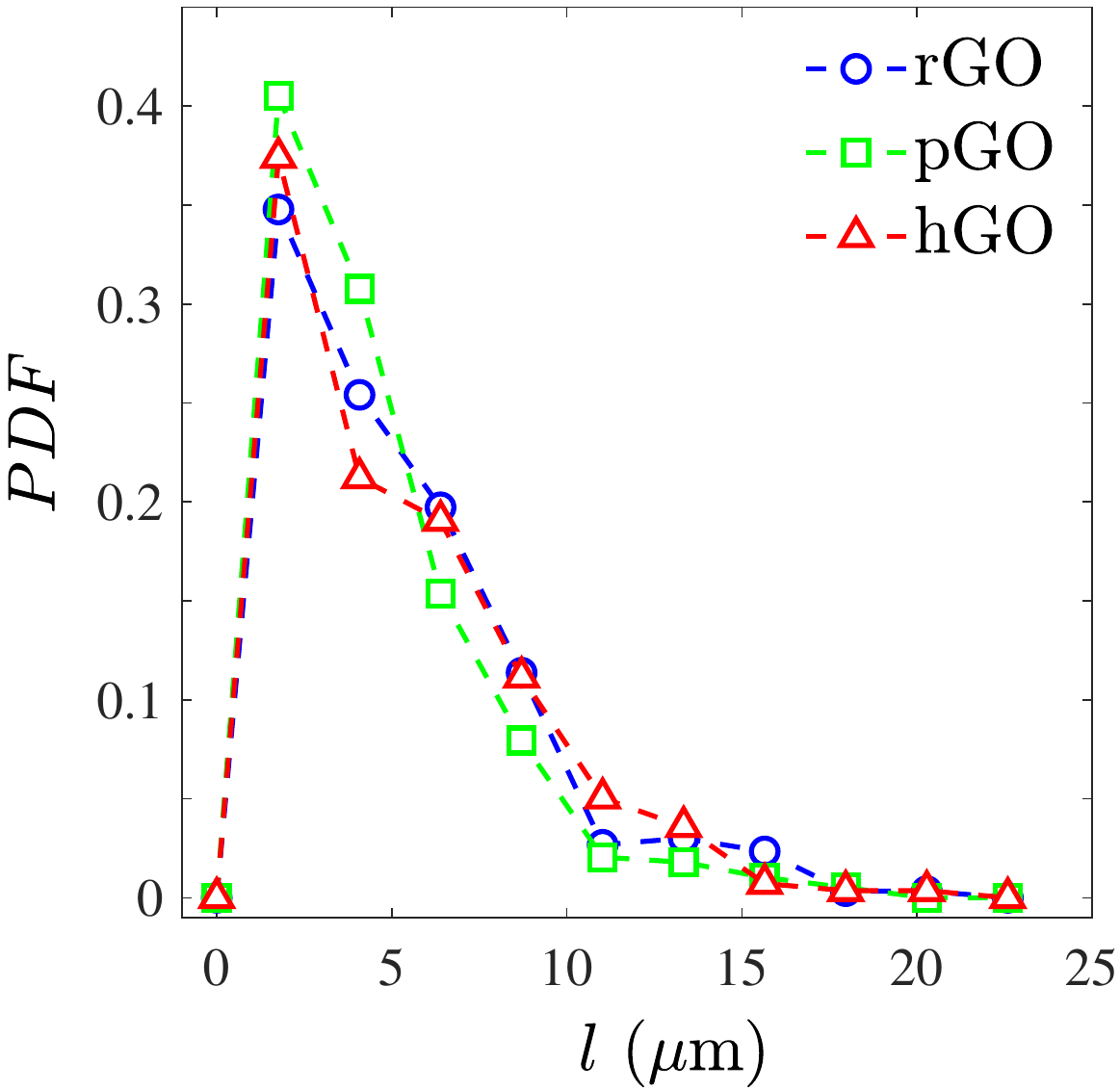} 
	\caption{The probability density functions of the utilized nanomaterials characteristic size. The PDFs are calculated from the coarse-resolution SEM imaging.}
	\label{fig:SEMsize}
\end{figure}

A vacuum-assisted synthesis method \cite{zhang2011vacuum} was used to reduce the oxygen content of hGO, generating pGO and rGO. Specifically, an Across International vacuum oven (AT09-UL) was first preheated to 120~$\mathrm{^oC}$. Then, hGO was heated in the oven for 2 hours generating pGO. Similarly, rGO was generated by heating hGO using the oven, but for 6 hours. Scanning Electron Microscopy (SEM) images were collected to characterize the size of rGO, pGO, and hGO. Both high-resolution (9.7~$\mathrm{nm/pixel}$ with the field-of-view of $14.9~\mu \mathrm{m} \times 9.9~\mu \mathrm{m}$) and coarse-resolution (0.5~$\mathrm{\mu m/pixel}$ with the field-of-view of $512.0~\mu \mathrm{m} \times 512.0~\mu \mathrm{m}$) were performed. For the high-resolution SEM, FEI Helios 650 nanolab was used, with the accelerating voltage and current being set to 5~kV and 25~pA, respectively. Such relatively small accelerating voltage facilitated \cite{michael2009challenges} capturing relatively detailed structure of the GO nanomaterials. Figures~\ref{fig:SEM}(a), (b), and (c) show exemplary high-resolution SEM images of rGO, pGO, and hGO, respectively. As evident in Fig.~\ref{fig:SEM}, all tested nanomaterials feature a layered structure similar to that presented in past studies, see for example~\cite{boateng2020significant,thiruppathi2020novel,mao2020efficient}. The insets of the images presented in Fig.~\ref{fig:SEM} show that the size of the stack of 2D layers is smaller than about 100~nm for all tested nanomaterials. Although the results presented in Fig.~\ref{fig:SEM} facilitate understanding the detailed structure of the utilized GO due to the high resolution of the imaging, they do not allow for measuring a characteristic length of the utilized particles. In order to facilitate this, large field-of-view/coarse-resolution SEM imaging was also performed for the GO flakes. For the coarse-resolution SEM imaging, Tescan Mira 3 XMU was used. Accelerating voltage and current were set to 20~kV and 15~pA, respectively. The coarse-resolution SEM images were binarized using MATLAB, and a characteristic length ($l$) was estimated for each detected particle using $l = \sqrt{4A_{\mathrm{p}}/\pi}$, with $A_{\mathrm{p}}$ being the area of a binarized particle image. For instance, for a circular disk, $l$ equates to the diameter of the disk. The probability density function of $l$ was estimated for all types of utilized nanomaterials, and the results are presented in Fig.~\ref{fig:SEMsize}. The bin size of the PDF distributions was set to 2.3 $\mu\mathrm{\mathrm{m}}$, and this was selected following a trial and error process, leading to the best presentation of the PDFs. All PDFs at $l = 0$ were set to zero. The results in the figure suggest that the PDF distribution is similar for all tested nanomaterials, with the maximum probability of about 2~$\mu \mathrm{m}$. Generally, the thermal behavior of the doped fuel is influenced by the nanomaterial type, the loading concentration, and the size distribution of the added nanomaterials. Since the PDF of the particle size is rather similar for all types of the utilized nanomaterials (see Fig.~\ref{fig:SEMsize}), the investigated combustion characteristics are expected to be influenced by the nanomaterial type and the loading concentration.\par

\begin{figure}[!t]
	\centering
	\includegraphics[width=0.5\textwidth]{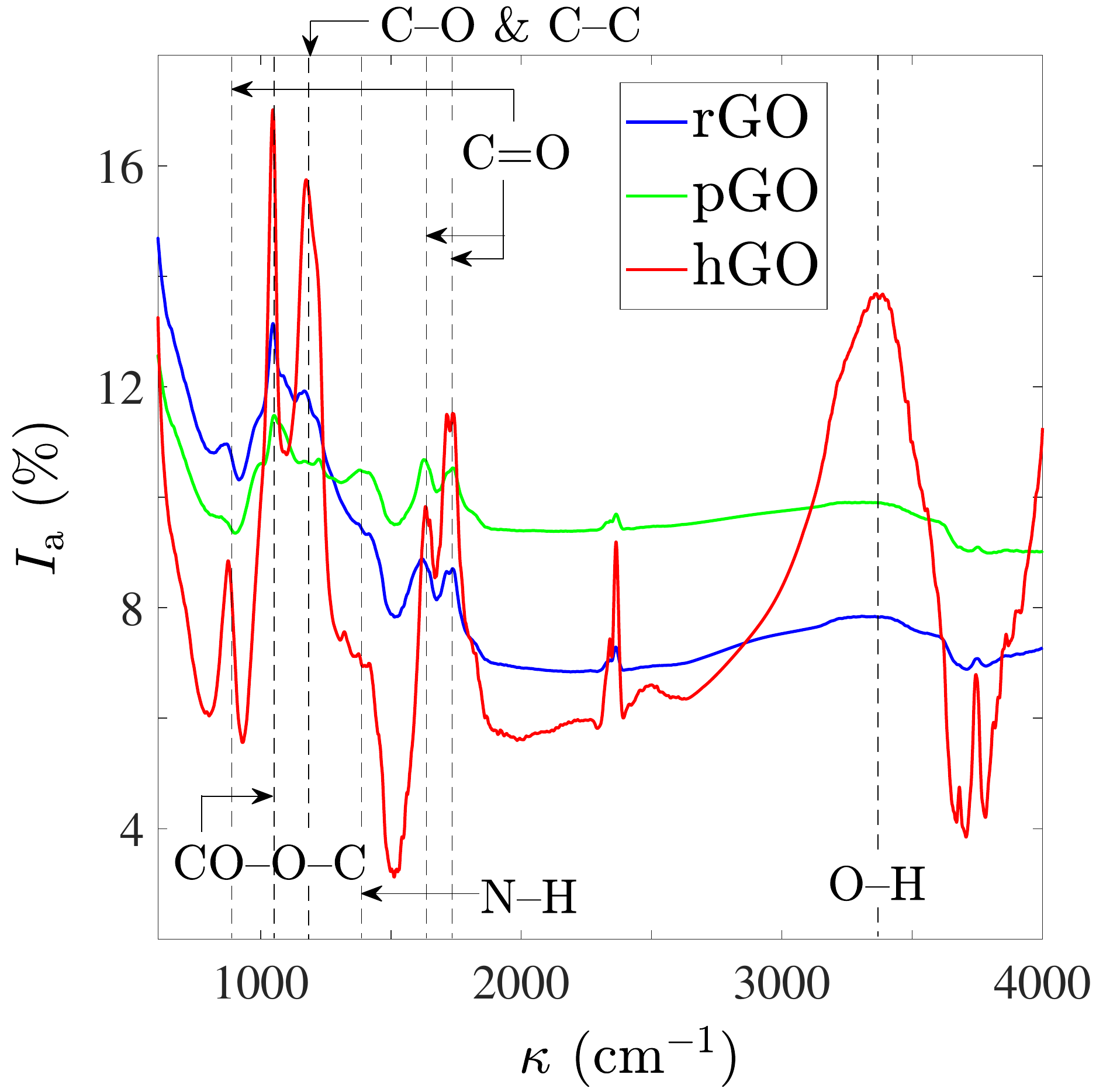} 
	\caption{Absorption intensity versus wave-number of the tested nanomaterials obtained from the FT-IR spectroscopy.}
	\label{fig:FT-IR}
\end{figure}

Although the results of the SEM imaging show that the synthesized nanomaterials feature relatively similar size distribution (for $l \gtrsim 2~\mu$m), further insight related to the chemical characteristics of the tested nanomaterials can be obtained using Fourier-Transform Infrared (FT-IR) spectroscopy \cite{zabihi2019simultaneous,poddar2018ultrasound}. The FT-IR experiments were performed using Thermo Scientific Nicolet iS20 Spectrometer, with the scanning range of 600 to 4000 $\mathrm{cm^{-1}}$ and a resolution of 0.48 $\mathrm{cm^{-1}}$. Figure~\ref{fig:FT-IR} presents the absorption intensity ($I_\mathrm{a}$) for rGO, pGO, and hGO, which are shown by the blue, green, and red colors, respectively. Absorption intensity is defined as the radiant power absorbed by the sample normalized by the radiant power of the incident light. Several dominant peaks are detected and related to the molecular structure of the tested nanomaterials. The results in the figure show that all tested materials feature peaks at 1050, 1110, 1240, and 3400 $\mathrm{cm^{-1}}$, which correspond to CO-O-C \cite{paluszkiewicz2001analysis}, C-O and C-C \cite{wood1996investigation,movasaghi2008fourier}, N-H \cite{el2005ftir}, and O-H \cite{dovbeshko2000ftir} groups, respectively. The peaks at 875, 1580, and 1720 $\mathrm{cm^{-1}}$ correspond to C=O \cite{schulz2007identification,fabian1995comparative,hospodarova2018characterization}. The peak at 2360 $\mathrm{cm^{-1}}$ is related to a background carbon dioxide gas and is not related to the molecular structure of the tested materials. The results in the figure show significant presence of carbon and oxygen bonds in the molecular structure of hGO compared to that of pGO and rGO. Thus, the material with absorption intensity variation highlighted by the red color should correspond to hGO. However, those shown by the blue and green colors feature similar spectra. \par

\begin{figure}[!t]
	\centering
	\includegraphics[width=1\textwidth]{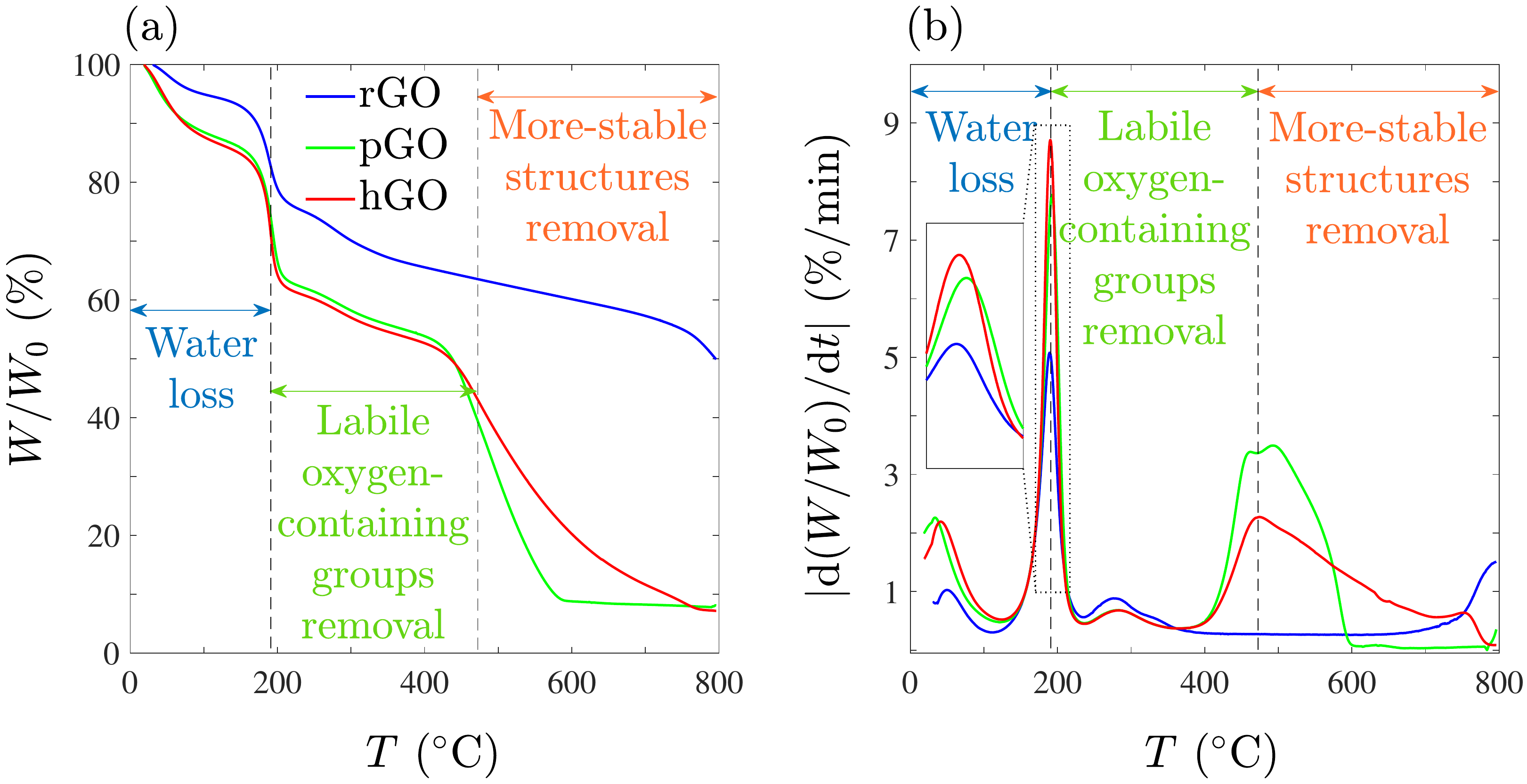} 
	\caption{The thermogravimetric analysis of the utilized nanomaterials. (a) $W/W_0$ versus temperature and (b) $|\mathrm{d}(W/W_0)/\mathrm{d}t|$ versus temperature.}
	\label{fig:TGA}
\end{figure}

The discussed method for preparation of the nanomaterials leads to hGO featuring more oxygen content than pGO, and similarly, pGO featuring more oxygen than rGO. The results presented in \mbox{Fig.~\ref{fig:FT-IR}} highlighted several (relatively) dominant peaks (related to oxygen-containing groups). However, the variations pertaining to rGO and pGO are rather similar. In order to investigate this, the Thermogravimetric Analysis (TGA) was performed using TGA Q500 V20.13, which operated with nitrogen gas. 5~mg batches of GO samples were used for the TGA. The temperature was varied from ambient to 800~$\mathrm{^oC}$, with an increase rate of 10~$\mathrm{^oC/min}$. Variations of the nanomaterials weight ($W$) normalized by that at room temperature ($W_0$) as well as $|\mathrm{d}(W/W_0)/\mathrm{d}t|$ versus temperature were obtained for all tested nanomaterials and are presented in \mbox{Figs.~\ref{fig:TGA}(a)} and (b), respectively. $W/W_0$ and $|\mathrm{d}(W/W_0)/\mathrm{d}t|$ related to rGO, pGO, and hGO are shown by the blue, green, and red curves, respectively. Three regions associated with significant decrease in the normalized materials weight can be identified in the figure. The region highlighted by the blue text pertains to the temperature ranging from the initial temperature of TGA samples to $\sim$190~$\mathrm{^oC}$ and corresponds to the evaporation of the adsorbed water~\mbox{\cite{shen2009fast,stankovich2007synthesis,wang2005synthesis}}. Studies of \mbox{\cite{marcano2010improved,shen2009fast,stankovich2007synthesis}} suggest that the weight loss pertaining to the temperature range of $\sim$190 and $\sim$470~$\mathrm{^oC}$ is due to the removal of unstable oxygen-containing functional groups, such as $\mathrm{CO}$, $\mathrm{COO}$, and $\mathrm{OH}$~\mbox{\cite{romero2018comparative}}. This region is highlighted by the green color text in \mbox{Figs.~\ref{fig:TGA}(a and b)}. The inset of \mbox{Fig.~\ref{fig:TGA}(b)} shows that, for $T\approx 190~\mathrm{^oC}$, the rate of weight loss pertaining to the red curve is larger than that of the green curve. Similarly, this parameter associated with the green curve is larger than that of the blue curve. This suggests that hGO features the largest oxidation level, followed by pGO and rGO. The region pertaining to $T \gtrsim 470~\mathrm{^oC}$ and highlighted by the orange text in \mbox{Figs.~\ref{fig:TGA}(a and b)} corresponds to decomposition of carbon from the skeleton of the nanomaterial~\mbox{\cite{wilson2009graphene}}. In essence, the results presented in \mbox{Fig.~\ref{fig:TGA}} suggest that the method of nanomaterials preparation, indeed, leads to synthesis (from hGO) of those with significant and relatively moderate loss of oxygen, which were referred to as rGO and pGO, respectively. \par

The synthesized nanomaterials were suspended in ethanol with 99\% chemical purity. Suspensions with three loading concentrations of 0, 0.01 and 0.1\% (by weight) were prepared. To avoid interference with the combustion experiments, no surfactant was used for stabilization during the suspension procedure. In fact, it has been shown that relatively stable suspensions of carbonaceous nanomaterials (for example graphite \cite{tanvir2016droplet,tanvir2017evaporation} and functionalized graphene sheets \cite{sabourin2009functionalized}) in ethanol can be prepared without using surfactants. In order to prepare the suspensions used in the present study, the GO powders were added to ethanol and sonicated using an ultrasound bath (Elma Ultrasonic Cleaner, EP100H) for 60 min prior to the experiments. To avoid the fuel temperature rise, the water inside the sonicator bath was replaced by room temperature water every 5 min. The prepared doped fuel remained uniform and stable for more than one hour. The combustion experiments were performed immediately after the completion of the sonication process. All tested conditions are tabulated in Table~\ref{tab:Conditions}. The first row pertains to a test condition without the addition of nanomaterials, which is pure ethanol. The first column presents the labels of the tested conditions. In this column, Er$X$, Ep$X$, and Eh$X$ refer to rGO, pGO, and hGO added to ethanol with loading concentration of $X$\%. The second column pertains to the type of the added nanomaterial to ethanol. The last column is the tested loading concentration, denoted by $\mathrm{[GO]\%}$. For the purposes of estimating the uncertainties of the reported results, each experiment was repeated 12 times.
\par

\begin{table} [h!]
	\centering
	\caption {The tested conditions.}
	\label{tab:Conditions}
	\renewcommand{\arraystretch}{1.2}
	\scalebox{1}{
		\begin{tabular}{ c  c  c }
			\hline
			\hline
			Tested Condition & Nanoparticle & $\mathrm{[GO]}$\% \\
			\hline
			Ethanol & --- 	& 0 		\\
			Er0.01 	& rGO	& 0.01 		\\
			Er0.1 	& rGO	& 0.1		\\
			Ep0.01  & pGO 	& 0.01		\\
			Ep0.1  	& pGO 	& 0.1		\\
			Eh0.01  & hGO 	& 0.01		\\
			Eh0.1   & hGO 	& 0.1		\\
			\hline
			\hline
	\end{tabular}}
\end{table}

\section{Experimental methodology}
\label{Methodology}

Details of the experimental setup, diagnostics, as well as the methods used for reduction of the collected data are presented in the following.  

\subsection{Experimental setup and diagnostics}
\label{Setup}

\begin{figure}[!t]
	\centering
	\includegraphics[width=1\textwidth]{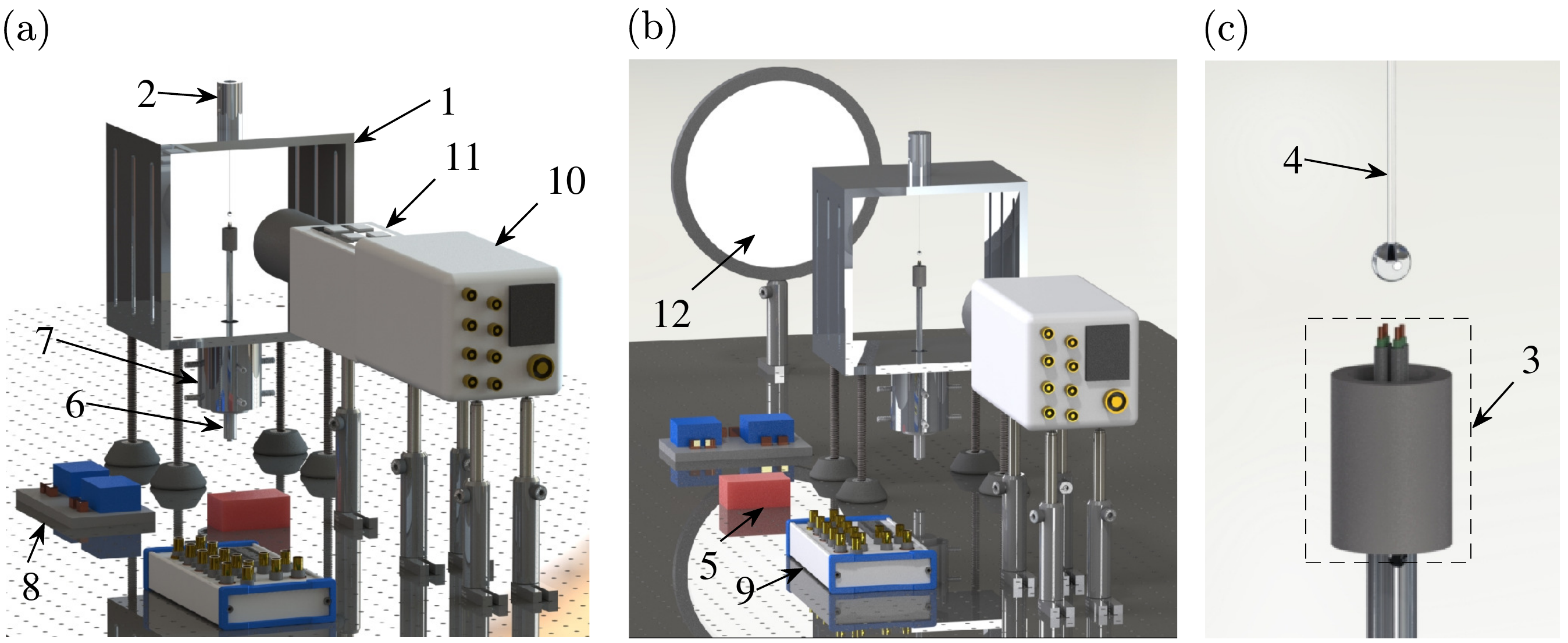} 
	\caption{Technical drawing of the experimental apparatus (a) $\mathrm{OH^*}$ chemiluminescence setup, (b) shadowgraphy setup, and (c) close view of the igniter and the droplet. The numbered items are: (1) setup holder box, (2) tube holder, (3) igniter module, (4) droplet suspension mechanism, (5) igniter power source, (6) solenoid actuator, (7) actuator holder, (8) relay module, (9) data acquisition unit (10) high-speed camera, (11) image intensifier, and (12) back-light LED.}
	\label{fig:Setup}
\end{figure}

\begin{figure}[!h]
	\centering
	\includegraphics[width=0.5\textwidth]{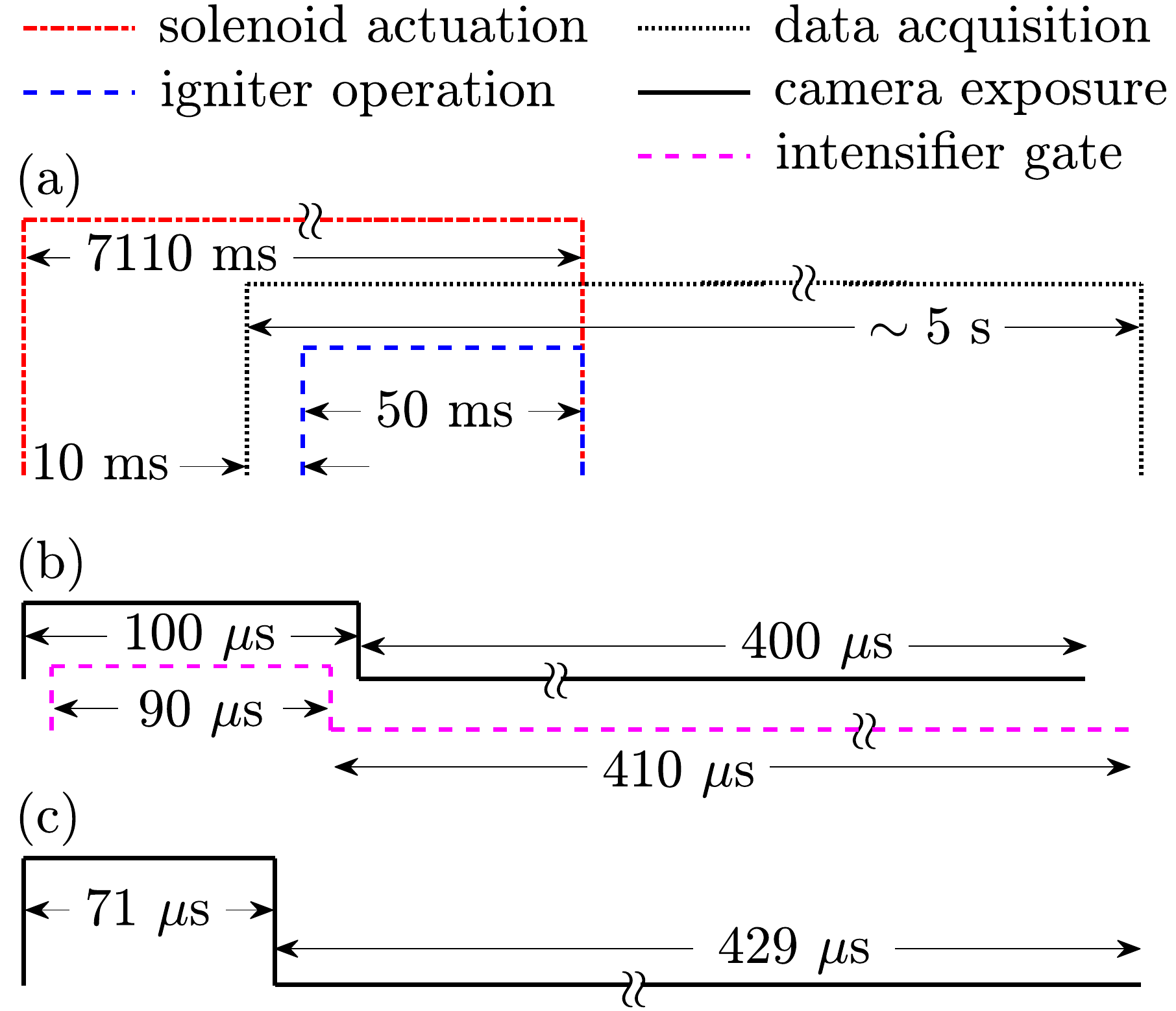} 
	\caption{(a) The relative timing of the solenoid actuation, igniter operation, and data acquisition. (b) The relative timing of the camera and intensifier for $\mathrm{OH^*}$ chemiluminescence imaging. (c) The timing of the camera for shadowgraphy imaging.}
	\label{fig:Signals}
\end{figure}

Separate chemiluminescence and shadowgraphy measurements were performed in the present investigation. Depending on the utilized measurement technique, two separate experimental arrangements were employed. Specifically, The experimental arrangements related to the chemiluminescence and shadowgraphy measurements are illustrated in Figs.~\ref{fig:Setup}(a) and (b), respectively. For clarity purposes, an inset of Figs.~\ref{fig:Setup}(a and b) is enlarged and shown in Fig.~\ref{fig:Setup}(c). In total, the experimental setups used for chemiluminescence and shadowgraphy experiments are composed of 12 components/modules, which are highlighted by the corresponding numbers in Fig.~\ref{fig:Setup}. A 20$\times$20$\times$15~$\mathrm{cm^3}$ aluminum box (item 1) that supports a tube holder at the top (item 2) and an igniter module at the bottom (item 3) were manufactured. Three grooves on the side walls of the box were machined. The grooves facilitate motion of a translation stage, which carries a calibration plate (not shown in the figure). Item 2 is an aluminum tube, screws into the box, carries 4 nylon set screws, and holds a droplet suspension tube (item 4). The tube is made of borosilicate, is 150~mm tall, has an outer diameter of 1~mm, and is referred to as the supporting mechanism in the present study. A propane torch was used to form a nearly symmetric bead with a diameter of $\mathrm{1.16}\pm\mathrm{0.01~mm}$ at the bottom of the tube, increasing the surface tension between the bottom of the tube and the liquid, and as a result, improving the droplet suspension. The droplet was deposited on the bead using a 100 $\mathrm{\mu L}$, model 1710, Gastight syringe from Harvard Apparatus. The droplet volume for all tested fuels is fixed, and the corresponding droplet diameter is $\mathrm{2.14}\pm\mathrm{0.01~mm}$. Smaller supporting mechanism diameter did not allow for proper suspension of the droplets. The igniter utilizes a plasma-assisted ignition strategy. Such ignition strategy is desired for droplet combustion research mainly due to the ability to control the duration of plasma generation and the relatively fixed output power, which improve the repeatability of the experiments \cite{bennewitz2020combustion}. The plasma arc duration is about 50~ms. The igniter module is composed of a four-prong (see Fig.~\ref{fig:Setup}(c)) plasma arc generator, an igniter control unit, a power source (item 5), a linear solenoid actuator (item 6), as well as an actuator holder (item 7). Eight set screws are installed on the holder, securing the actuator. Care was taken to position the igniter module components with respect to the droplet such that the igniter arc is positioned about 1~mm below the droplet when the actuator reaches its full range of motion. Items 5 and 6 are controlled using an electric relay (item 8) and an NI PCIe 6361 data acquisition board (item 9). LabVIEW was used to control the board and the diagnostics, which are discussed in the following. The relative timing of the solenoid actuation, igniter operation, and data acquisition signals are presented in Fig.~\ref{fig:Signals}(a). As shown in the figure, the data acquisition (dotted black curve) begins 60~ms prior to return of the actuator (see the descend in the red dotted dashed curve); and, the igniter operates 10 ms after the start of the data acquisition (see the blue dashed curve) for 50~ms. \par

High-speed $\mathrm{OH^*}$ chemiluminescence and shadowgraphy techniques were separately used in the experiments. The former and the latter techniques are employed to measure the ignition delay, similar to the studies of~\cite{mbugua2020ignition,marchese2011ignition} and the droplet burning rate, similar to the studies of~\cite{ooi2016graphite,pandey2019high,javed2013evaporation,javed2014effects,sim2018effects,ghamari2017combustion}, respectively. For both chemiluminescence and shadowgraphy techniques, a high-speed camera (Photron Fastcam Nova S12), shown as item 10 in Fig.~\ref{fig:Setup}, was used. The imaging acquisition frequency for both techniques was set to 2000~Hz. For $\mathrm{OH^*}$ chemiluminescence measurements, a Nikkor 50~mm ($f/\# = 1.4$) lens was used to connect an Invisible Vision UVi 1850B intensifier (item 11) to the camera. The intensifier was equipped with a UV Nikon Rayfact 105 mm ($f/\# = 1.4$) lens. A $\mathrm{310}\pm\mathrm{10~nm}$ band-pass filter was installed on the UV lens to collect flame chemiluminescence near wavelength of the $\mathrm{OH^*}$ emission. As illustrated in Fig.~\ref{fig:Signals}(b), the camera exposure time was set to 100~$\mu$s. The intensifier gate was set to 90~$\mu$s, and it was centered with respect to the camera exposure time. The intensifier gain was set to 45\%. The pixel resolution of the chemiluminescence measurements (defined as the ratio of the field-of-view side length to the corresponding number of pixels) is $\mathrm{44~\mu m/pixel}$. For the shadowgraphy experiments, the camera was equipped with a Zeiss Makro-Planar 100~mm (f/\#=8) lens. As schematically illustrated in Fig.~\ref{fig:Signals}(c), the camera exposure time was $\mathrm{1/14000~s\approx71~\mu s}$. A FotodioX C-300RS FlapJack LED (item 12) was used for background illumination. The LED brightness was obtained using a trial and error process, maximizing the quality of the acquired images. The LED brightness was fixed for all tested conditions. The pixel resolution (with similar definition provided for the chemiluminescence imaging) of the shadowgraphy technique was $\mathrm{38~\mu m/pixel}$.
\par

\subsection{Data reduction}
\label{Datareduction}

\begin{figure}[!t]
	\centering
	\includegraphics[width=0.5\textwidth]{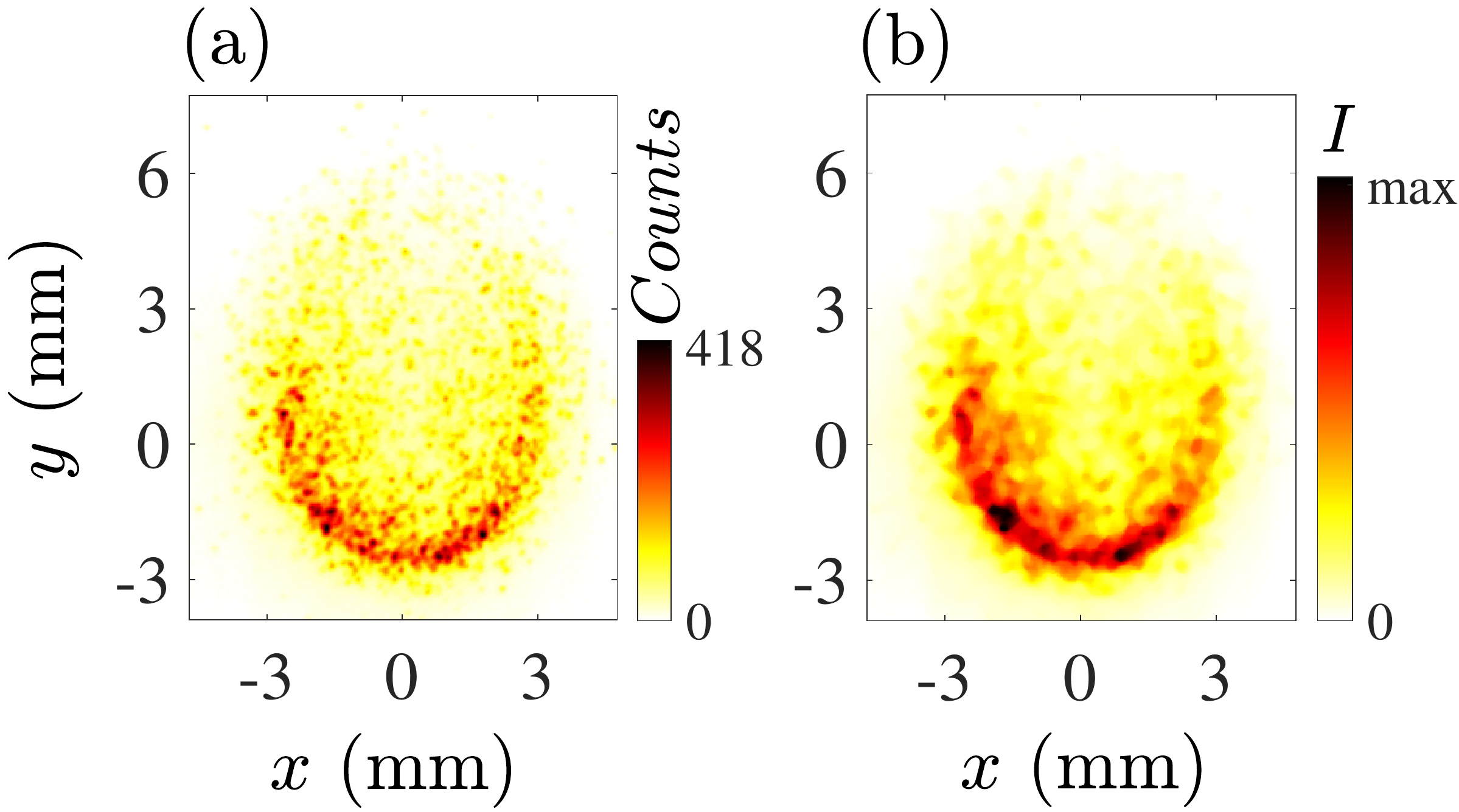} 
	\caption{Chemiluminescence data reduction: (a) raw chemiluminescence image and (b) after background subtraction, white field correction, and median-based filtering.}
	\label{fig:OHdatareduction}
\end{figure}

In this subsection, the procedures for reduction of the chemiluminescence and shadowgraphy images, which are used for studying the ignition delay and the droplet burning rate, are discussed. The chemiluminescence data reduction is similar to that discussed in several past investigations, see for example~\cite{heydarlaki2020experimental,heydarlaki2021competing,mollahoseini121flame}. Figure~\ref{fig:OHdatareduction}(a) presents a raw chemiluminescence image. All acquired chemiluminescence images were first subtracted by a background. The background was obtained by collecting and averaging 1000 images with the camera lens capped. Then, using intensifier settings identical to that used for the chemiluminescence measurements, 1000 white field images were collected and averaged using the LED light, shown in Fig.~\ref{fig:Setup}. As discussed in past investigations \cite{heydarlaki2020experimental}, the position of the LED does not influence the white field images. For all collected white field images, the LED brightness was set to 30\%. Then, the background-subtracted chemiluminescence images were divided by the white field to normalize for the spatially dependent sensitivity of the imaging equipment. In order to minimize electronics noise, a 7$\times$7 $\mathrm{pixels}^2$ window median-based filter was applied to the white field-corrected chemiluminescence images. The resultant filtered image is shown in Fig.~\ref{fig:OHdatareduction}(b). It was confirmed that reducing the size of the utilized filter to 3$\times$3 $\mathrm{pixels}^2$ windows did not influence the reported results significantly. \par

\begin{figure}[!t]
	\centering
	\includegraphics[width=1\textwidth]{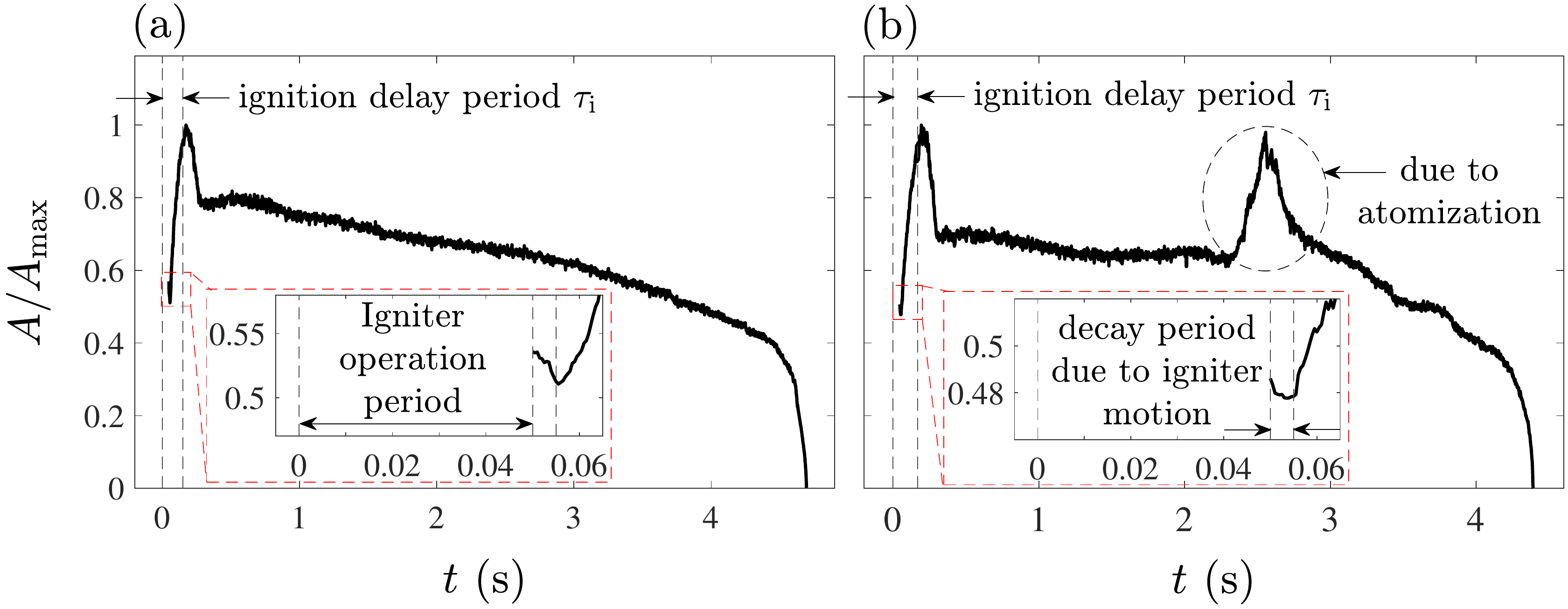} 
	\caption{Variation of $A/A_\mathrm{max}$ versus time for the test conditions of (a) Ep0.01 and (b) Eh0.1.}
	\label{fig:OHcurve}
\end{figure}

It can be inferred from the study of Aggarwal~\cite{aggarwal2014single} that the ignition delay ($\tau_{\mathrm{i}}$) is the time difference between the instant that the droplet is introduced to the ignition source and the instant that the flame size reaches its maximum value. Similarly, Bennewitz et al.~\cite{bennewitz2019systematic,bennewitz2020combustion} calculated the ignition delay using the flame area measured by infrared imaging. A similar approach is used in the present study, except the $\mathrm{OH^*}$ chemiluminescence images are used compared to the infrared images. Here, the summation of the area of the pixels that feature $\mathrm{OH^*}$ chemiluminescence larger than zero was obtained and is referred to as $A$. Variation of $A/A_\mathrm{max}$ for two representative conditions of Ep0.01 and Er0.01 are shown in Figs.~\ref{fig:OHcurve}(a) and (b), respectively. It was verified that using larger threshold values does not significantly change variation of $A$ normalized by the corresponding maximum value ($A_\mathrm{max}$) versus time. In Figs.~\ref{fig:OHcurve}(a) and (b), $t = 0$ corresponds to the instant that the igniter spark is formed. This was facilitated by starting the data acquisition 10~ms prior to igniter operation, as shown by the dotted black and dashed blue curves in Fig.~\ref{fig:Signals}(a). As schematically demonstrated in the figure, the igniter spark operates for about 50~ms. During $0 \lesssim t \lesssim 50$~ms, the chemiluminescence signal is contaminated by the light emission from the igniter spark, and as a result, the corresponding data is presented for $t \gtrsim 50$~ms, see the inset of Fig.~\ref{fig:OHcurve}(a). After $t \approx 50$~ms, the solenoid is retracted during $50 \lesssim t \lesssim 55$~ms. This was confirmed through high-speed imaging of the igniter retraction process, which is not presented here for brevity. During the solenoid retraction time period ($50 \lesssim t \lesssim 55$~ms), the flame size decreases, causing the decreasing values of $A/A_\mathrm{max}$, see the insets of Fig.~\ref{fig:OHcurve}(a and b). Such behavior is independent of the tested conditions, and is speculated to be linked to the motion of the igniter. Nevertheless, as discussed in Appendix A, the igniter spark and its retracting motion only influence the flame dynamics for a short period of time ($\sim$5~ms), which is relatively small compared to the droplet lifetime.

After $t\gtrsim 55$~ms, $A/A_\mathrm{max}$ increases and features a global maximum, see Figs.~\ref{fig:OHcurve}(a and b). Ignition delay ($\tau_\mathrm{i}$) is defined as the time difference between the instant that the spark is formed (i.e. $t=0$) and the instant that the normalized flame area reaches 0.95 of its global maximum, see this duration highlighted in Figs.~\ref{fig:OHcurve}(a and b). Analysis of the results presented in Fig.~\ref{fig:OHcurve} suggests that $A/A_\mathrm{max}$ features relatively large fluctuations near the global maximum. This leads to large uncertainty in estimation of the time at which $A/A_\mathrm{max}$ equals unity. In order to reduce this uncertainty, 95\% of $A/A_\mathrm{max}$ was used for estimating the ignition delay.

\begin{figure}[!t]
	\centering
	\includegraphics[width=0.5\textwidth]{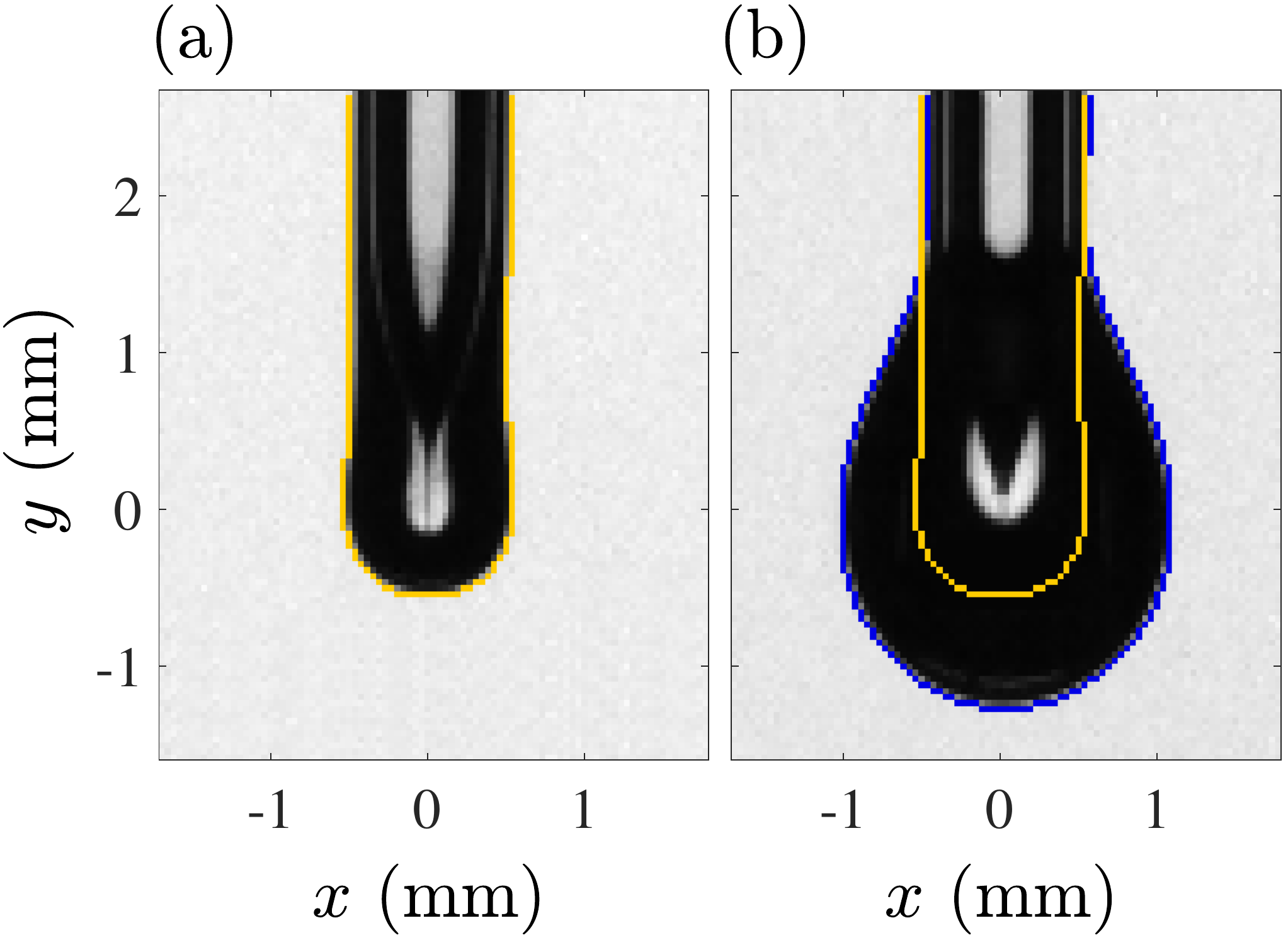} 
	\caption{Shadowgraphy image obtained for (a) the supporting mechanism and (b) the droplet suspended on the supporting mechanism.}
	\label{fig:SGdatareduction}
\end{figure}

To estimate the droplet equivalent diameter (which is used for burning rate estimation), first, the image of the utilized droplet supporting mechanism was acquired. This image is shown in Fig.~\ref{fig:SGdatareduction}(a). Then, using an edge finding algorithm in MATLAB, the exterior of the supporting mechanism was obtained, see the yellow curve shown in Fig.~\ref{fig:SGdatareduction}(a), and overlaid on Fig.~\ref{fig:SGdatareduction}(b). The same algorithm was also used to estimate the edge of the droplet, which is shown by the blue curve in Fig.~\ref{fig:SGdatareduction}(b). Then, the droplet and the supporting mechanism volume were calculated. In order to do this, first, the combined droplet and supporting mechanism volume was assumed to be made of several disks, stacked along the vertical axis. The volume of each disk is $\pi r(y)^2 \Delta y $, where $\Delta y = 38~\mu$m is the pixel height and $r(y)$ is the radius of the disk at a given height of $y$. Then, the total droplet and supporting mechanism volume ($V$) was calculated by summing up the disks volumes. A similar approach was used to calculate the volume of the supporting mechanism ($V_\mathrm{s}$) as well. The droplet equivalent diameter $D$ was then obtained, using $D = [6(V-V_{\mathrm{s}})/\pi]^{1/3}$.
\par

\begin{figure}[!t]
	\centering
	\includegraphics[width=0.5\textwidth]{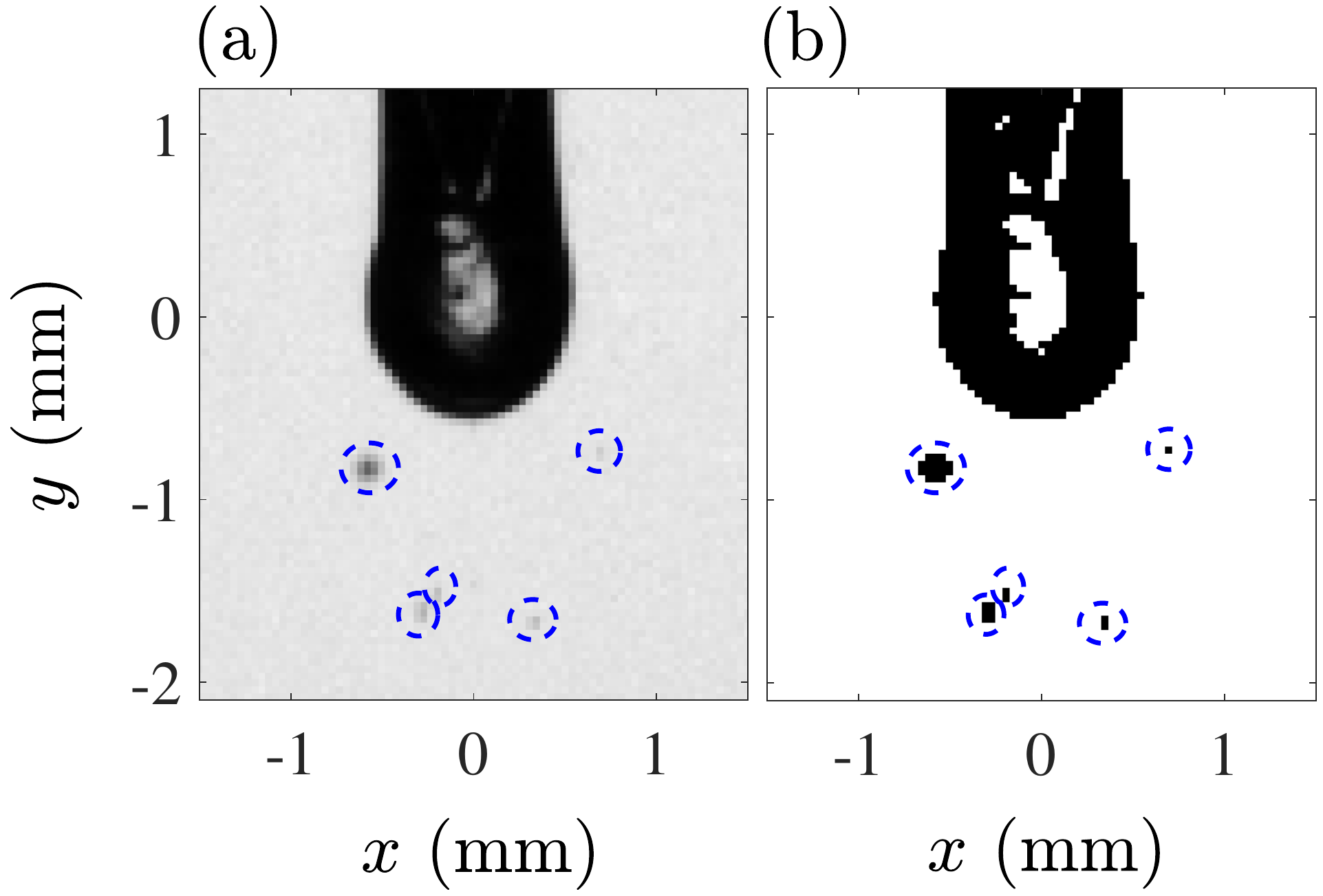} 
	\caption{(a) A representative raw shadowgraphy image and (b) the binarized image using the adaptive thresholding technique. The baby droplets are circled in both (a) and (b).}
	\label{fig:Ejectiondata}
\end{figure}

The projected trajectory of the secondary droplets and their mass were obtained using the shadowgraphy images. This is rather challenging since the ejected droplets move out of the camera's focusing volume during their lifetime as well as presence of background noise. Binarization with a spatially constant threshold removes some of the baby droplets from the images, which is undesired. To address this, an adaptive threshold \cite{bradley2007adaptive} technique was utilized. To demonstrate the performance of this technique, a raw shadowgraphy image is shown in Fig.~\ref{fig:Ejectiondata}(a), and the detected baby droplets after binarization of the raw image with the adaptive thresholding method is presented in Fig.~\ref{fig:Ejectiondata}(b). Care was taken to ensure no secondary droplet is removed as a result of the applied technique, and that the baby droplets are distinguished from the background noise. This requires estimation of the overall imaging resolution and its implementation for the baby droplets detection, with detailed discussions provided in Appendix B.

\section{Results}
\label{results}

The results are grouped into three subsections. Those related to the ignition delay, characterization of atomization, and contribution of atomization to the burning rate are presented in subsections~\ref{ignitiondelay}, \ref{atomchar}, and \ref{atomcont}, respectively. 

\subsection{Ignition delay}
\label{ignitiondelay}

Following the procedure discussed in section~\ref{Datareduction}, the ignition delay was estimated. $\tau_{\mathrm{i}}$ pertaining to pure ethanol related to the present study is shown in Table~\ref{tab:ID}. Also, presented in the table are the ignition delays for pure ethanol droplets extracted from the studies of \cite{bennewitz2020combustion,ghamari2016experimental}. As can be seen, the ignition delay varies from 71 to 274~ms. Several factors could alter the ignition delay. The ignition method, the location of the igniter with respect to the droplet, the droplet supporting mechanism, and the droplet size could cause the differences between the ignition delays reported in Table~\ref{tab:ID} for pure ethanol. Correcting for the effects of the igniter and supporting mechanism on the ignition delay is not possible. However, $\tau_{\mathrm{i}}$ can be normalized for the effect of the droplet size on the ignition delay, using the droplet diameter. Since heat transfer by convection and buoyancy are, respectively, related to the droplet surface area and volume, $\tau_{\mathrm{i}}$ was also normalized by $D_0^2$ and $D_0^3$. As the results in Table~\ref{tab:ID} suggest, such normalization does not allow for the collapse of the data, and the scatter of the ignition delay results for ethanol is expected to be due to the differences between the utilized igniter, supporting mechanism, and droplet size in the present study and those of \cite{bennewitz2020combustion,ghamari2016experimental}.
\par

\begin{table} [!t]
	\centering
	\caption {Ethanol droplet ignition delay of the present study and those reported in \cite{bennewitz2020combustion} and \cite{ghamari2016experimental}.}
	\label{tab:ID}
	\renewcommand{\arraystretch}{1.2}
	\scalebox{1}{
		\begin{tabular}{ c c c c c }
			\hline
			\hline
			Study & $D_0$~(mm) & $\tau_\mathrm{i}$~(ms) & $\tau_\mathrm{i} /D^2_0~\mathrm{(ms/mm^2)}$ & $\tau_\mathrm{i} /D^3_0~\mathrm{(ms/mm^3)}$  \\
			\hline
			Present investigation 									& 2.14 	& 158.5	& 34.6  & 16.2  \\
			Bennewitz et al.~\cite{bennewitz2020combustion} & 1.35	& 71  	& 38.9  & 28.5  \\
			Ghamari~\cite{ghamari2016experimental}	  		& 0.79 	& 274	& 439.0 & 555.7 \\
			\hline
			\hline
		\end{tabular}}
\end{table}

\begin{figure}[!t]
	\centering
	\includegraphics[width=0.45\textwidth]{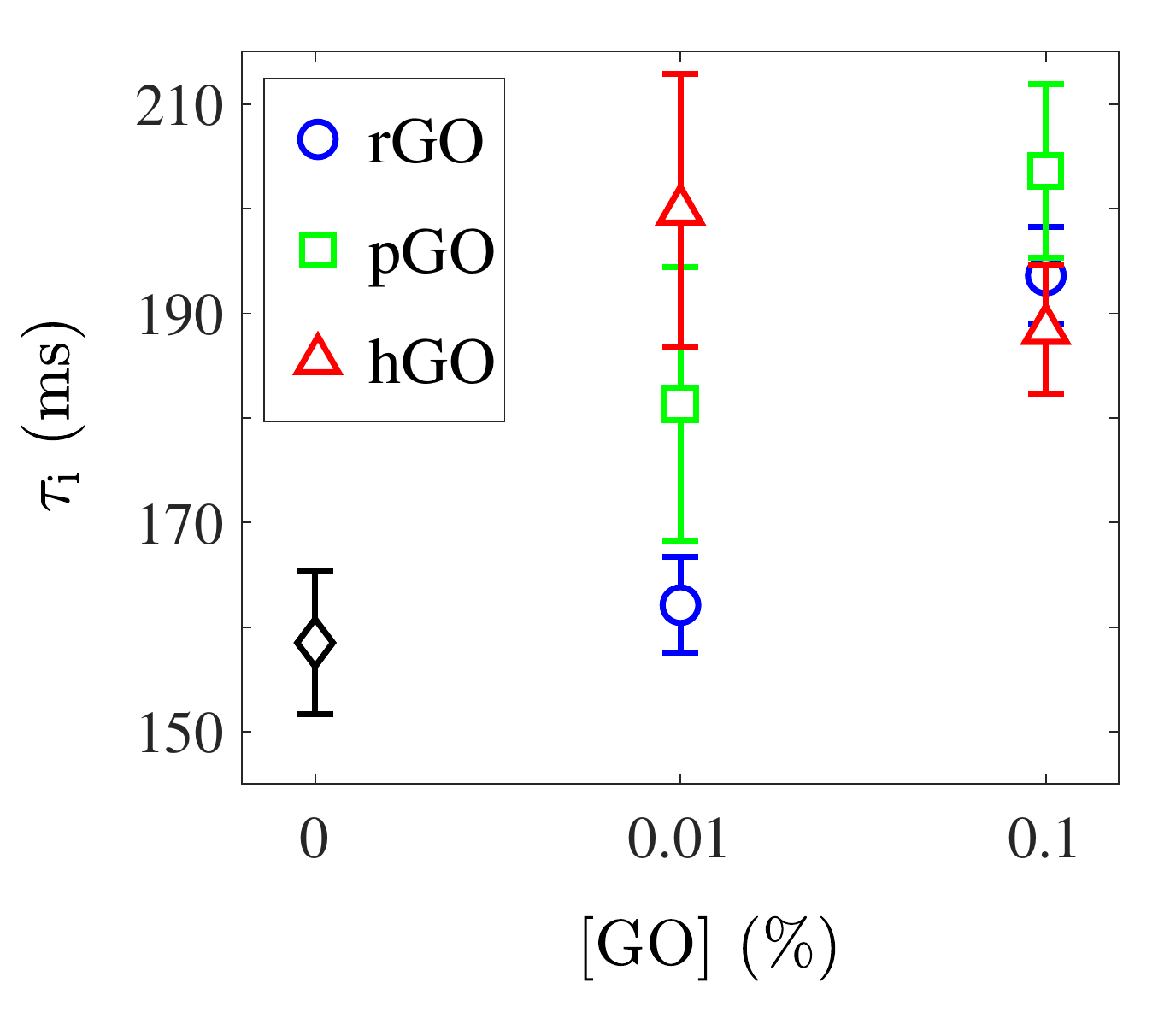} 
	\caption{Effect of graphene oxide type and concentration on the ignition delay of ethanol droplets.}
	\label{fig:IgDelay}
\end{figure} 

Nevertheless, since the igniter, the suspension mechanisms, and the droplet diameter are unchanged in the present investigation, comparisons for the effect of the nanomaterial addition on the droplet ignition delay can be made. The ignition delay results are presented in Fig.~\ref{fig:IgDelay}. In the figure, the error bars pertain to the repeatability of the experiments. Specifically, the length of the error bar equals the difference between the maximum and minimum of the estimated ignition delay for 12 repeats of each test condition. The results in Fig.~\ref{fig:IgDelay} show that, for ethanol doped with rGO and pGO, the ignition delay increases almost linearly with increasing the loading concentration. However, for hGO, the effect of the loading concentration on the ignition delay is different. That is, for hGO, maximum ignition delay pertains to the loading concentration of 0.01\%. The results in the figure show that for loading concentration of 0.01\%, a maximum increase of 18.3, 9.2, and 6.8\% occurs for the ignition delay of ethanol droplets doped with hGO, pGO, and rGO, respectively. The maximum increase of the ignition delay for the loading concentration of 0.1\% are 20.9, 16.8, and 16.6\% pertaining to ethanol droplets doped with pGO, rGO, and hGO, respectively. The increase in the ignition delay of liquid fuels doped with carbonaceous nanomaterials is consistent with the observations reported in~\cite{singh2019effect,singh2020effect,bennewitz2020combustion}. It is speculated that, generally, the presence of the carbonaceous nanomaterials increases the absorption of the heat generated by the flame without an immediate increase in the droplet temperature. Thus, at a given time and prior to the ignition completion (flame area reaching 0.95 of its maximum), the mass fraction of vapor formed around the droplet surface is expected to be smaller for the doped ethanol compared to pure ethanol droplet. As a result, it takes longer for the doped droplet to ignite (compared to pure ethanol). Nonetheless, the effect of GO addition on ethanol ignition delay is presented here for the first time. \par

\begin{figure}[!h]
	\centering
	\includegraphics[width=0.5\textwidth]{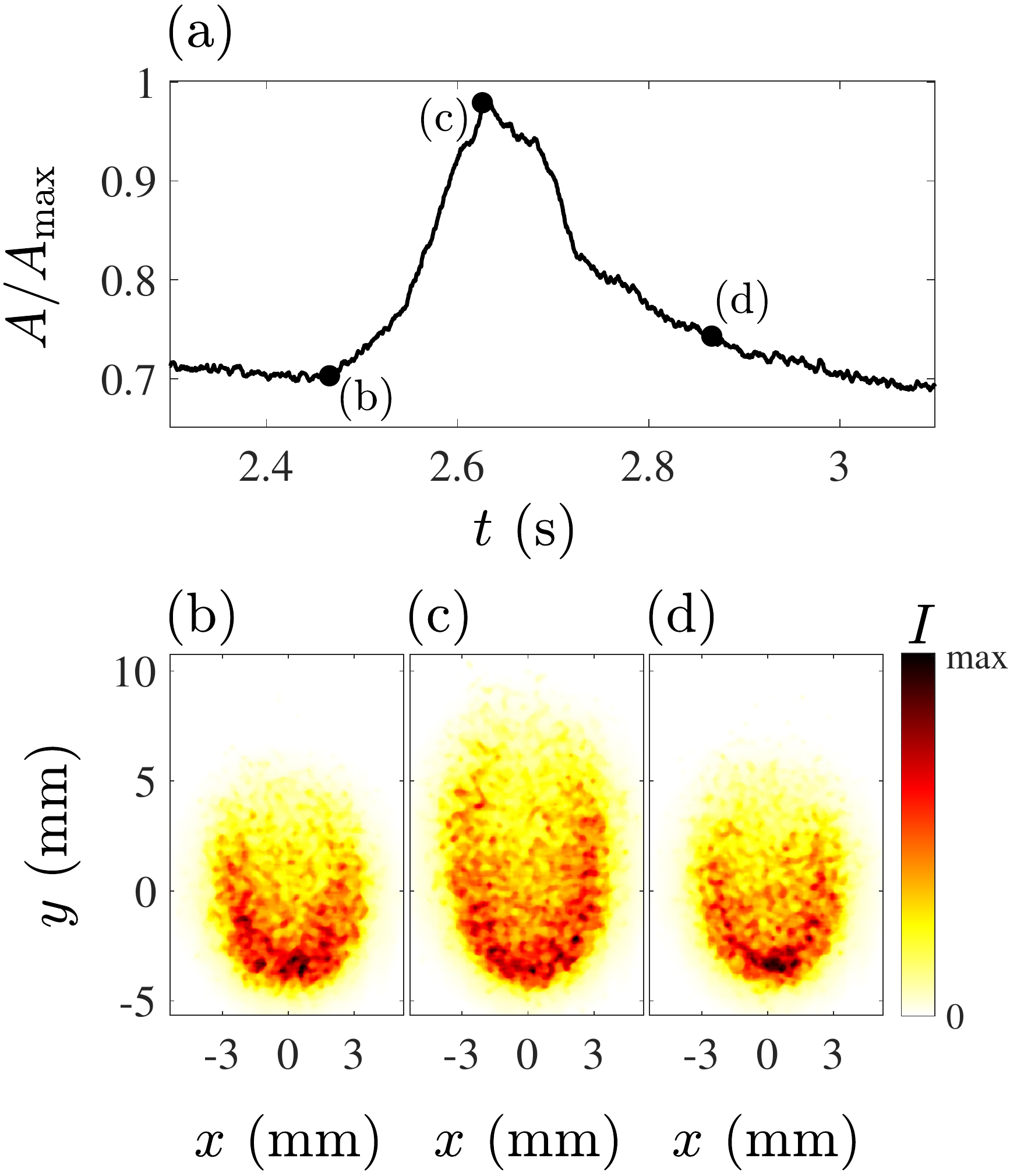} 
	\caption{(a) Variation of the normalized flame area versus time and (b-d) are the $\mathrm{OH^*}$ chemiluminescence images pertaining to $t = 2.47$, 2.63, and 2.87~s, respectively. The results pertain to Ep0.01 test condition.}
	\label{fig:Atomicpeak}
\end{figure}

Although the chemiluminescence measurements allow for characterizing the ignition delay, further insight regarding the droplet burning dynamics can be obtained using such measurements. Results presented in Fig.~\ref{fig:OHcurve}(a) show that, after the ignition delay period, $A/A_\mathrm{max}$ decreases with time. Our results show that, for some test conditions, the reduction of the droplet surface area is followed by a significant increase in the flame surface area, see the second peak in Fig.~\ref{fig:OHcurve}(b). Variation of $A/A_\mathrm{max}$ versus time pertaining to the second peak and for the test condition of Ep0.01 is shown in Fig.~\ref{fig:Atomicpeak}(a). The flame chemiluminescence images pertaining to $t = 2.47$, 2.63, and 2.87~s, are shown in Figs.~\ref{fig:Atomicpeak}(b), (c), and (d), respectively. As can be seen, at $t = 2.63$, the flame surface area is maximized, which is mainly due to the increase of the flame size along the vertical axis. Similar observations are reported in the literature, see for example~\cite{miglani2014insight}. Past studies, such as Basu et al.~\cite{basu2016combustion}, argue the reason for such behavior can be linked to the droplet atomization. Specifically, during the atomization period, ejection of the baby droplets can increase the flame size, as demonstrated in Fig.~\ref{fig:Atomicpeak}(c). The ejected baby droplets cannot be visualized by the $\mathrm{OH^*}$ chemiluminescence; however, further details related to the atomization characteristics are studied using the shadowgraphy technique and are discussed in the following subsections. After the atomization, the flame surface area decreases until the flame extinction takes place.
\par

\subsection{Atomization characterization}
\label{atomchar}

\begin{figure}[!t]
	\centering
	\includegraphics[width=1\textwidth]{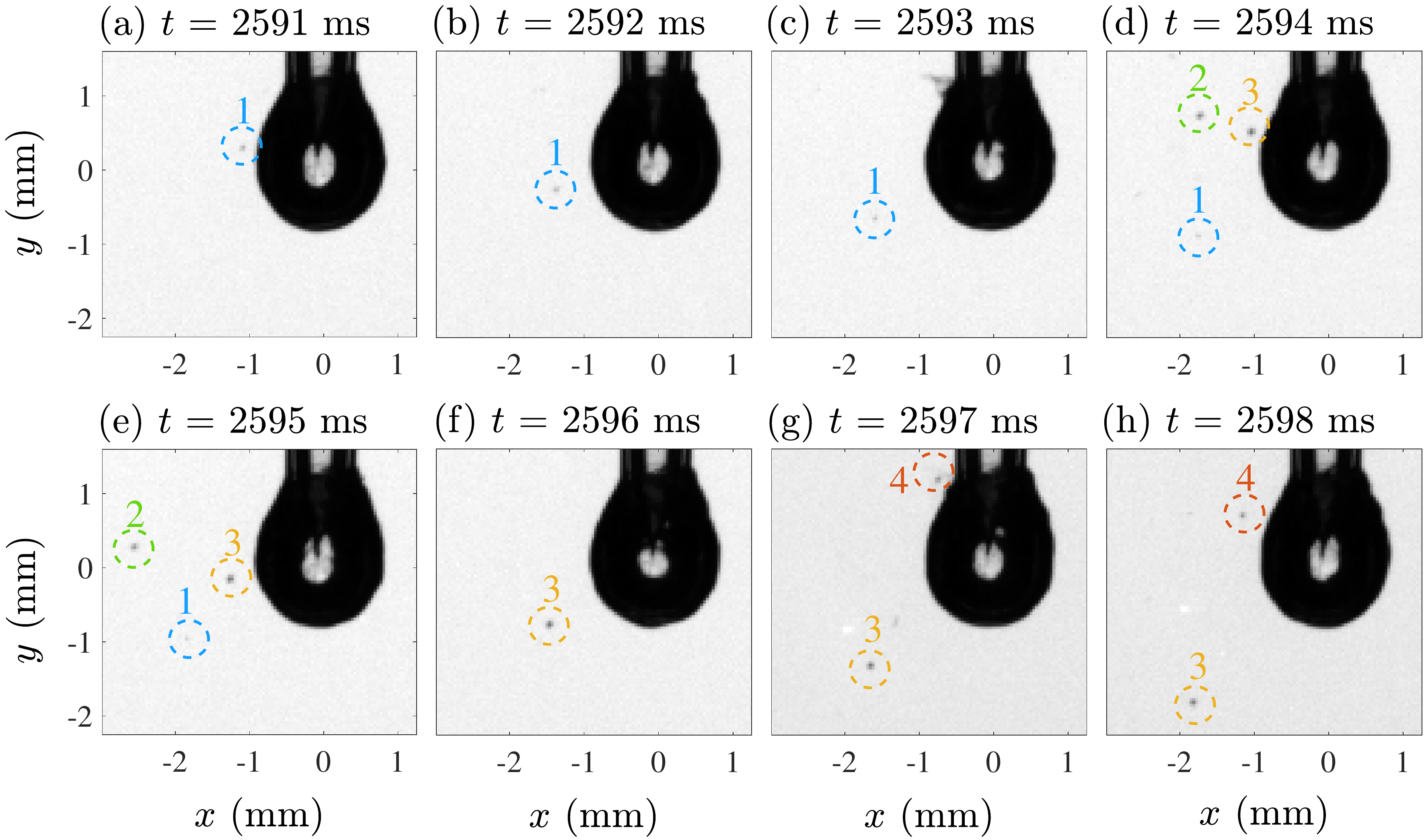} 
	\caption{Shadowgraphy images of baby droplets formation and ejection. The results pertain to the test condition of Eh0.01.}
	\label{fig:Ejectionevent}
\end{figure}

Figures~\ref{fig:Ejectionevent}(a--h) show a series of shadowgraphy images of a burning droplet during the atomization period. The images are separated by 1~ms corresponding to $2591~\mathrm{ms}~\leq~t~\leq~2598~\mathrm{ms}$, and pertain to test condition of Eh0.01. For the test condition presented Figs.~\ref{fig:Ejectionevent}(a--h), the atomization period started at $t = 2521~\mathrm{ms}$ and finished at $t = 3075~\mathrm{ms}$. Thus, only exemplary results are presented in the figure. As can be seen, four baby droplets (highlighted by \#1--4) are formed and ejected from the main droplet. The baby droplets are color-coded based on the first instant that they are observed. For example, baby droplet \#1 is observed at $t=2591~\mathrm{ms}$ and exists for about 4~ms, as shown in figures~\ref{fig:Ejectionevent}(a--e). Analysis of all shadowgraphy images suggests that atomization may occur for all tested conditions of doped ethanol droplets. However, pure ethanol droplets do not feature atomization. Past studies, see for example~\cite{basu2016combustion,wang2019experimental,miglani2014insight,gan2012combustion}, suggest that the difference between the boiling points of molecules inside a multi-component fuel with one another and/or with water (due to potential water vapor condensation for alcohols \cite{lee1992experimental}), the existence of a droplet supporting mechanism, and/or presence of solid particle doping agents can lead to the formation of bubble(s) inside the droplet. Since doping agents are not present for our pure ethanol experiments, and that atomization is not observed for these experiments, it is not expected that the presence of the supporting mechanism or the difference between the boiling points of ethanol and possibly condensed water cause the atomization. Thus, it is speculated that the presence of GO nanomaterials is the potential reason for bubble formation, and as a result, the atomization.
\par

\begin{figure}[!h]
	\centering
	\includegraphics[width=0.6\textwidth]{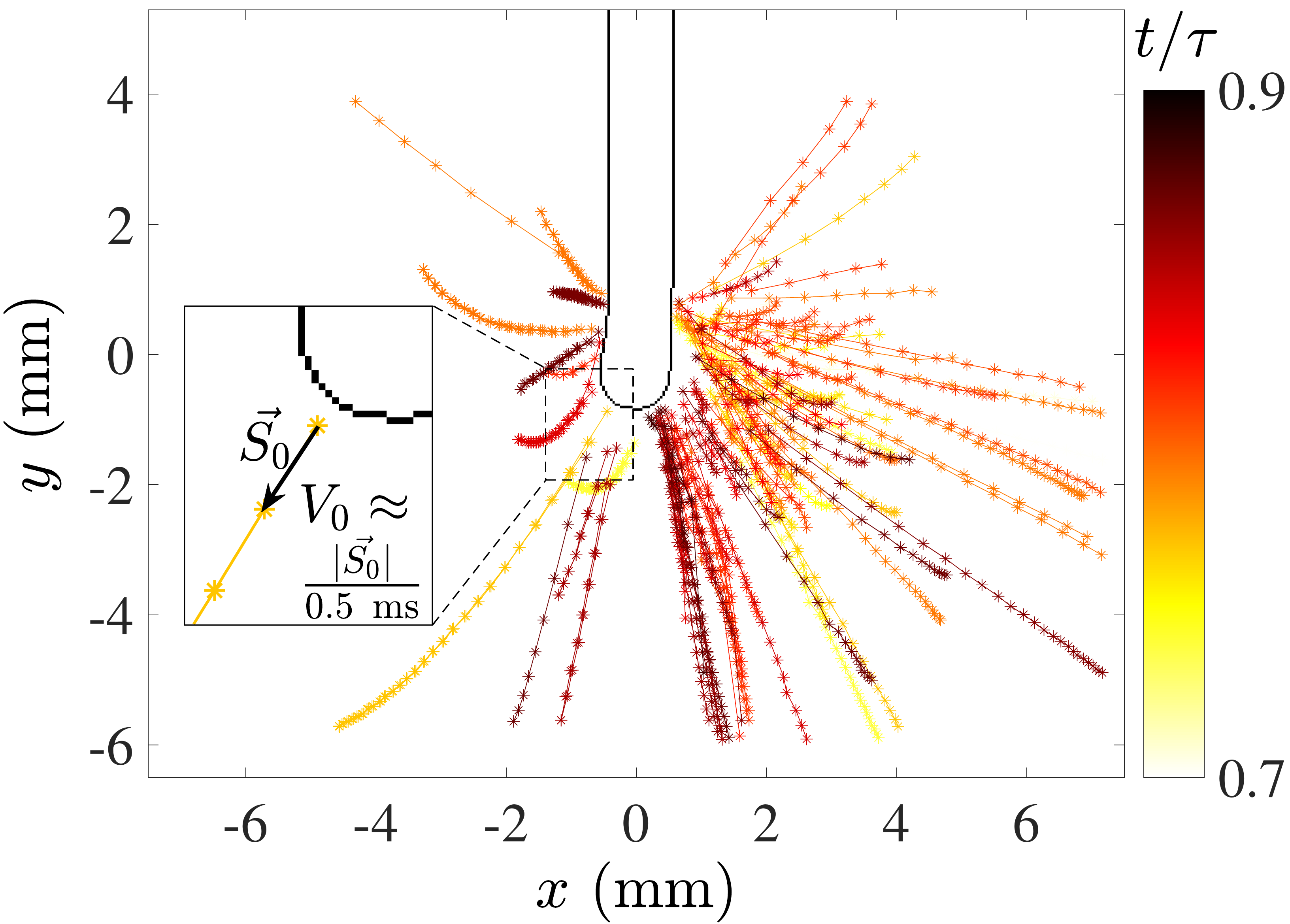} 
	\caption{Baby droplet ejection projected trajectories. The results pertain to the tested condition of Er0.01.}
	\label{fig:Ejectioncapture}
\end{figure}

For all tested conditions, the baby droplets were identified following the procedure discussed in section~\ref{Datareduction}. Figure~\ref{fig:Ejectioncapture} presents the projected trajectory of all baby droplets corresponding to the test condition of Er0.01. The supporting mechanism is marked by the black solid curve in the figure. The trajectories are color-coded based on the time at which the corresponding baby droplet appears first. In the figure, $t$ is normalized by the total droplet lifetime ($\tau$). As can be seen, all the ejection events take place during $t/\tau \geq 0.7$. It can also be observed that the location at which a baby droplet is initially formed and ejected is rather random. For example, the dark red trajectories (occurring towards the end of the droplet lifetime) may appear on both sides of the droplet. \par

\begin{figure}[!t]
	\centering
	\includegraphics[width=1\textwidth]{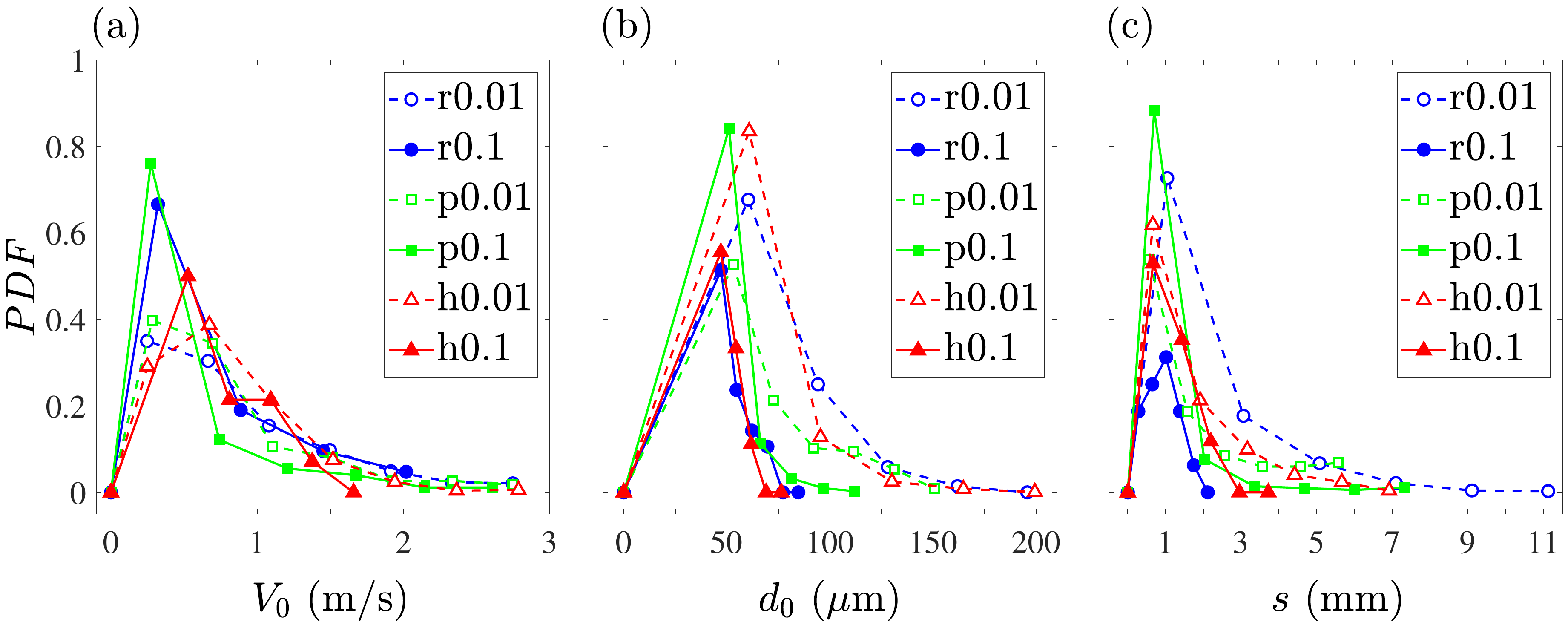} 
	\caption{The probability density functions of (a) the $x-y$ projected initial velocity of the ejections, (b) the initial diameter of the ejected baby droplets, and (c) the $x-y$ projected trajectory length of the ejections.}
	\label{fig:PDFs}
\end{figure}

For all tested conditions, the magnitude of the initial $x-y$ projected velocity of the baby droplets ($V_0$) were estimated using the corresponding trajectory and the finite differencing scheme \cite{lomax2013fundamentals}. Specifically, the magnitude of the projected baby droplet initial displacement ($\vec{S_0}$) was obtained and normalized by the corresponding time period ($\mathrm{0.5~ms}$). This is schematically shown in the inset of Fig.~\ref{fig:Ejectioncapture}. The probability density function of $V_0$, the initial $x-y$ plane projected baby droplet diameter ($d_0$), and the total length of an $x-y$ plane projected trajectory ($s$) were obtained for all tested conditions. The PDFs of $V_0$, $d_0$, and $s$ are presented in Figs.~\ref{fig:PDFs}(a), (b), and (c), respectively. In the figure, open and solid symbols correspond to 0.01 and 0.1\% by weight concentrations of the doping agent, respectively. Blue, green, and red colors pertain to rGO, pGO, and hGO, respectively. The resolution of the PDFs depends on the number of the data points available due to the baby droplet ejections and spatio-temporal resolution of the measurements. The bin size for the PDFs was selected using a trial and error process, which leads to the best presentation of the results. All PDFs at zero values of $V_0$, $d_0$, and $s$ were set to zero. The results show that the most probable value of $V_0$ is 0.5$\pm$0.2~m/s, which is comparable to the ethanol and air laminar flame speed at the stoichiometric condition. The results in Fig.~\ref{fig:PDFs}(b) show that the most probable value of the ejected baby droplet initial diameter corresponds to $40~\mu \mathrm{m} \leq d_0 \leq 60~\mu \mathrm{m}$, with the majority of the test conditions featuring a most probable $d_0$ value of about 50~$\mu$m. This is about 55 times smaller than the main droplet equivalent initial diameter ($D_0$). The values of $V_0$ and $d_0$ estimated in the present study are similar to those reported in~\mbox{\cite{yao2021atomization}}. Specifically, Yao et al.~\mbox{\cite{yao2021atomization}} showed that, for a 2~mm diameter ethanol droplet undergoing microexplosions, the baby droplets with the diameter of $\sim 50~\mu$m feature a mean projected velocity of $\sim 0.5$ m/s. The most probable value of $s$ is about 1~mm, as shown in Fig.~\ref{fig:PDFs}(c), which is comparable to the main droplet equivalent diameter. The results in Fig.~\ref{fig:PDFs}(c) show that the possibility of relatively long projected trajectories occurrence is non-zero, similar to those shown in Fig.~\ref{fig:Ejectioncapture}. \par

\begin{figure}[!t]
	\centering
	\includegraphics[width=1\textwidth]{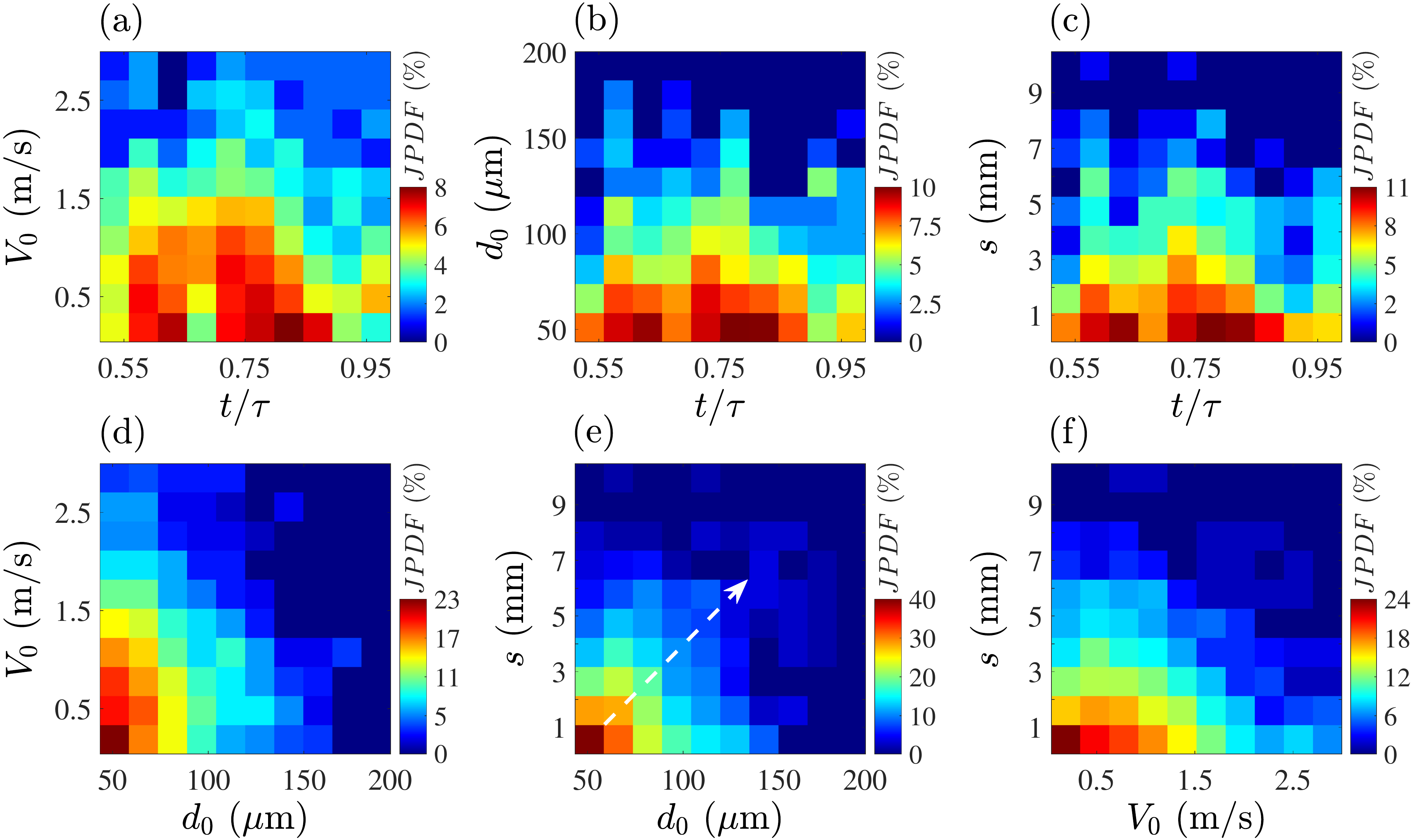} 
	\caption{JPDF of (a) the initial projected velocity of the baby droplets, (b) the initial diameter of the ejected baby droplets, and (c) the length of the ejected baby droplets projected trajectories versus time. (d), (e), and (f) are the JPDFs of $V_0$ versus $d_0$, $s$ versus $d_0$, and $s$ versus $V_0$, respectively.}
	\label{fig:JPDFs}
\end{figure}

Although the results presented in Fig.~\ref{fig:PDFs} allow for understanding the most probable values of $V_0$, $d_0$, and $s$, they do not demonstrate (potential) correlations between the most probable values. The Joint Probability Density Function (JPDF) of the above parameters was calculated to address this. Our analysis shows, generally, the time duration (compared to the droplet lifetime) and the number of occurrences of the baby droplet ejections are relatively small for each tested condition, which does not facilitate a proper presentation of the JPDFs. However, the results presented in Fig.~\ref{fig:PDFs} showed that the PDFs of the baby droplet ejections are not significantly influenced by the amount of doping as well as the type of the added nanomaterials. Thus, data related to all tested conditions pertaining to the atomization were combined, facilitating the calculation of the JPDFs with improved resolution of the presentation. The JPDF of $V_0$, $d_0$, and $s$ versus time are presented in Figs.~\ref{fig:JPDFs}(a--c), respectively. The results in the figure show that the most probable value of the above parameters occurs around $ 0.55 \lesssim t/\tau \lesssim 0.95$, which is in agreement with the results presented in Fig.~\ref{fig:Ejectioncapture}. Figures~\ref{fig:JPDFs} (d--f) present the JPDFs of $V_0$, $d_0$, and $s$ versus one another. The results in Fig.~\ref{fig:JPDFs}(d) show that relatively large baby droplets may be ejected with relatively large initial velocities. Also, relatively large baby droplets may travel longer trajectories, see the white dashed arrow in Fig.~\ref{fig:JPDFs}(e). However, the results in Figs.~\ref{fig:PDFs}~and~\ref{fig:JPDFs}(d--f) show that majority of the ejected baby droplets are relatively small, feature projected velocity that is comparable to the flame speed, and travel relatively short distances compared to the droplet equivalent diameter. This suggests the majority of the ejected baby droplets feature a relatively short lifetime and they may burn prior to leaving the flame. However, there exist relatively large baby droplets with large projected trajectory length, which may escape the flame envelope. Nonetheless, assessing this requires simultaneous shadowgraphy and flame chemiluminescence measurements, which is not performed in the present study. \par

\subsection{Contribution of atomization to doped ethanol burning rate}
\label{atomcont}

\begin{figure}[!t]
	\centering
	\includegraphics[width=0.5\textwidth]{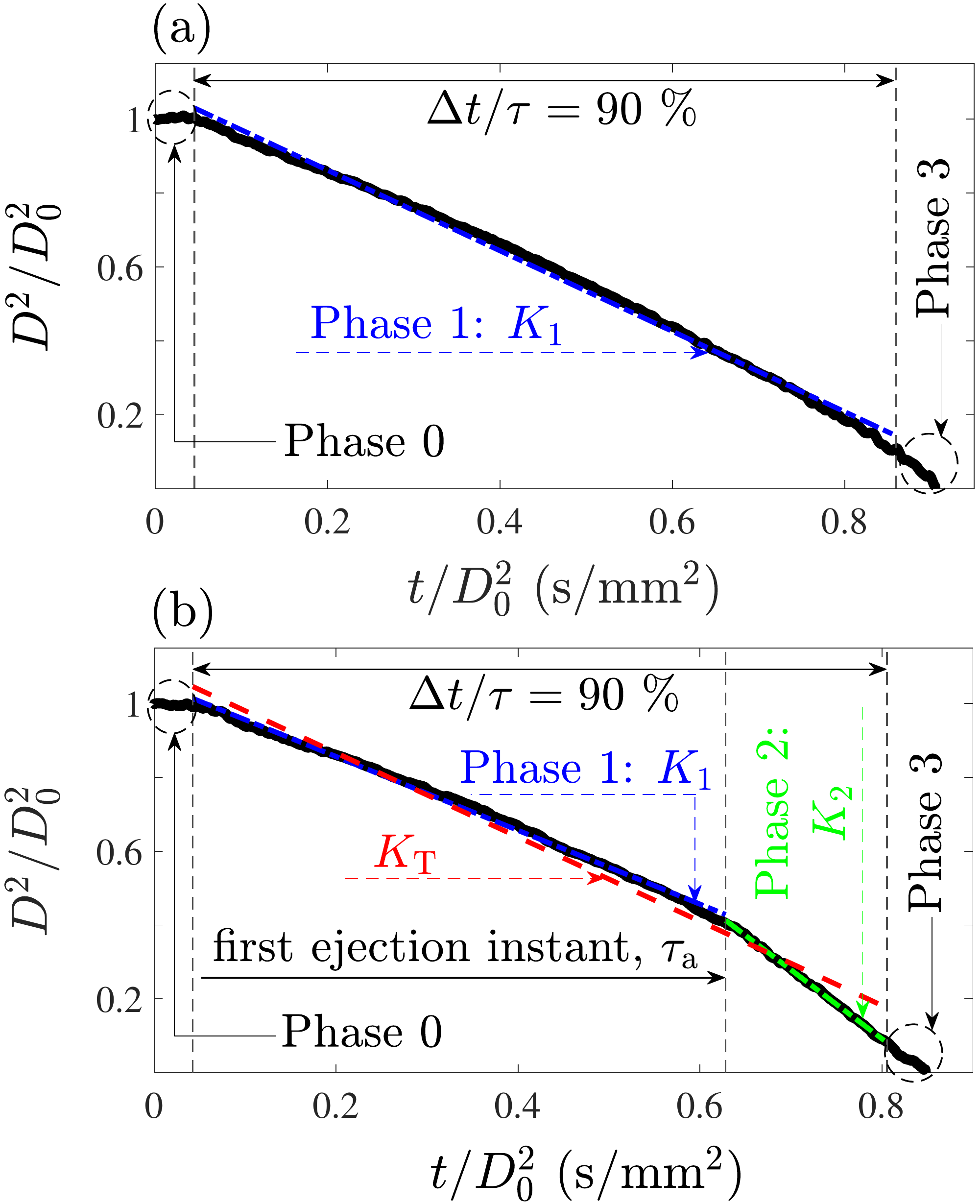} 
	\caption{Regression of the droplet equivalent diameter squared versus time (a) without atomization and (b) with atomization. The results in (a) and (b) pertain to Er0.1 and Ep0.01 test conditions, respectively. Both the vertical and horizontal axes values are normalized by the droplet initial equivalent diameter squared ($D^2_0$).}
	\label{fig:SGcurve}
\end{figure}

Although the above discussion allows for characterizing the baby droplets ejections, the effect of such ejections on the burning rate is unknown and is investigated in this subsection. Variation of the droplet equivalent diameter squared versus time for two test conditions with and without atomizations are presented in Figs.~\ref{fig:SGcurve}(a) and (b), respectively. The results are normalized by the droplet initial equivalent diameter squared ($D^{\mathrm{2}}_{\mathrm{0}}$). In the figures, $t = 0$ corresponds to the instant that the igniter spark is initially formed. The results in Fig.~\ref{fig:SGcurve} show that, for about $0.05~(\mathrm{s/mm^2})\times 2.14^2~(\mathrm{mm^2}) \approx 0.2~\mathrm{s}$, the droplet equivalent diameter does not change significantly. During this period, highlighted as Phase 0 in Figs.~\ref{fig:SGcurve} (a and b), the droplet size is influenced by both gasification (during which the combustible mixture is formed) and the droplet thermal expansion. The former reduces the droplet diameter; however, the latter increases the diameter. As can be seen, the effects of these two competing mechanisms on the droplet equivalent diameter cancel out, leading to a nearly constant value of the droplet equivalent diameter during Phase 0. This observation is expected and is consistent with the results reported in the literature for pure ethanol~\cite{ghamari2016experimental}, $\alpha$-methylnaphthalene~\cite{nishiwaki1955kinetics}, and Bakken crude oil \cite{singh2020effect}.
\par

Analysis of all test conditions with and without atomization suggests that about 5\% of the droplet lifetime corresponds to Phase 0. This was also confirmed by calculating the change in the slope of the $D^2/D_0^2$ variation versus time. After the heat-up period (Phase 0), for the test conditions without atomization, the droplet equivalent diameter squared decreases linearly with a nearly constant slope. Following the regression analysis presented in Appendix C, it was obtained that about 90\% of the total droplet lifetime leads to the most accurate estimation of the slope. This period is highlighted in Figs.~\ref{fig:SGcurve}(a and b) as $\Delta t/\tau=90\%$. The rest of the droplet lifetime ($0.95 \lesssim t/\tau \lesssim 1$) pertains to the flame extinction period, and is referred to as Phase 3. \par

Compared to the results presented in Fig.~\ref{fig:SGcurve}(a), those shown Fig.~\ref{fig:SGcurve}(b) suggest that the diameter squared regression curve of the droplet features a noticeable change in the slope. In Fig.~\ref{fig:SGcurve}(b), the time instant at which the first baby droplet is ejected is highlighted by the dashed line at $t/\tau = 0.63$. As can be seen, this time instant corresponds to when the slope of the curve changes. Similar observations were made for the rest of the tested conditions that feature atomization. For the test condition presented in Fig.~\ref{fig:SGcurve}(b), the period between the end of Phase 0 and the instant that the first baby droplet is ejected is referred to as Phase 1. The atomization period is highlighted by Phase 2. For the test conditions without the baby droplet ejections, Phase 1 is limited between the end of Phase 0 and beginning of Phase 3. For test conditions with no atomization, Phase 2 does not exist.
\par

Past studies~\cite{turns1996introduction,bennewitz2019systematic,basu2016combustion,sim2018effects} suggest that the non-linear behavior of the $D^2$ regression curves may occur due to several reasons, which are non-spherical shape of the droplets~\cite{turns1996introduction}, the utilized liquid fuels being made of multi-components~\cite{bennewitz2019systematic}, atomization or micro-explosions ~\cite{bennewitz2019systematic,sim2018effects,basu2016combustion}, as well as the presence of the suspension equipment \cite{bennewitz2019systematic}. In the present investigation, during Phase 1 and for all tested conditions, the droplets are non-spherical and the suspension mechanism exists. Yet, the droplet equivalent diameter squared features a linear variation with a constant slope during this phase. For this reason and similar to past studies~\cite{ghamari2017combustion,pandey2019high,sim2018effects,bennewitz2020combustion}, a line was fit to the surface regression curves pertaining to Phase 1 for all tested conditions. The absolute value of the slope of the line pertaining to Phase 1 is referred to as $K_1$. In the present study, the slope of the regression curve changes once atomization takes place, but the slope remains nearly constant after the occurrence of the atomization and the slope absolute value is referred to as $K_2$. \par

\begin{figure}[!h]
	\centering
	\includegraphics[width=0.35\textwidth]{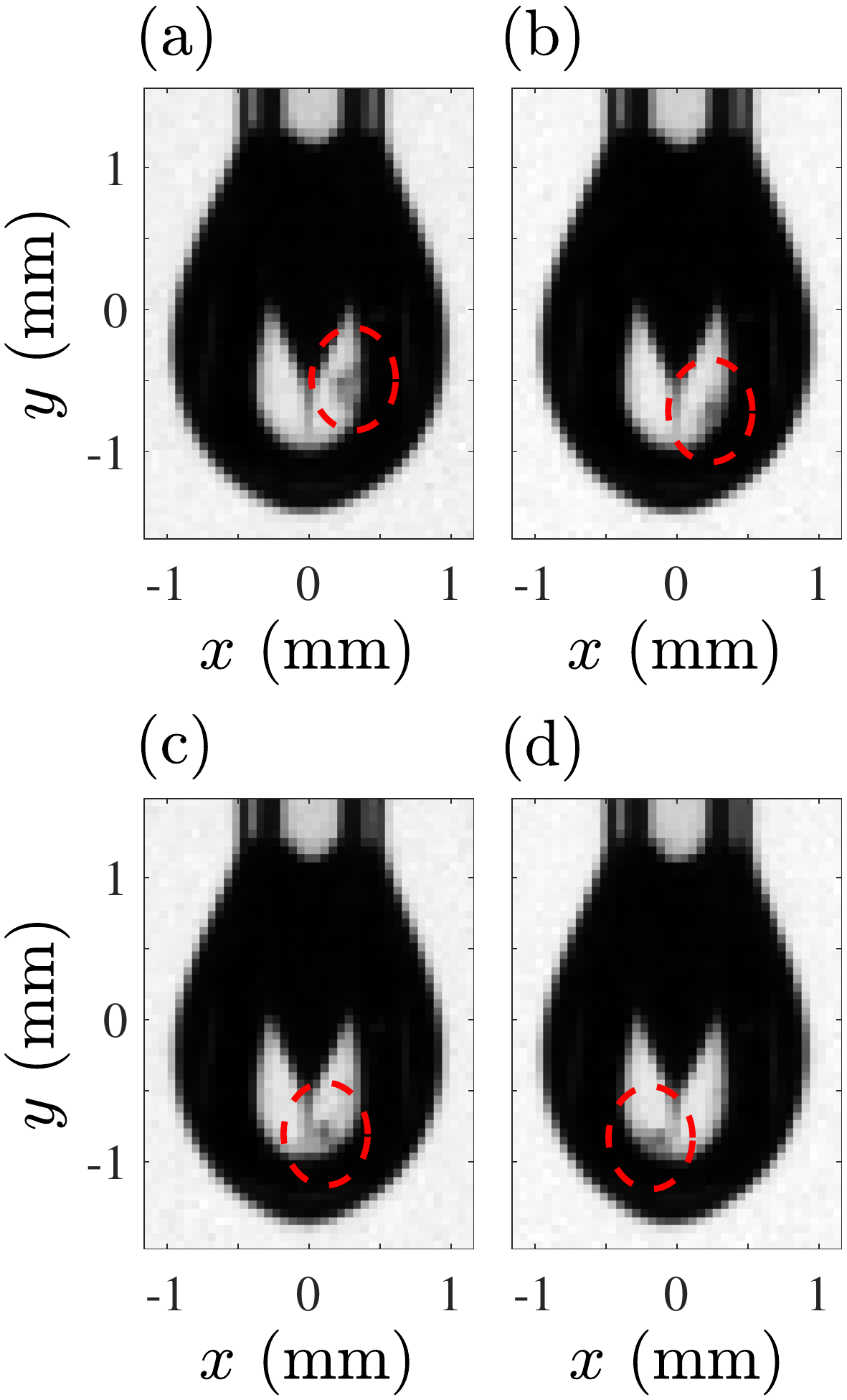} 
	\caption{Shadowgraphy images of ethanol droplet doped with 0.01\% of rGO during combustion. Images in (a--d) are separated by 75 ms and pertain to $t = 2200$, 2275, 2350, and 2425~ms. The red dashed circle highlights the location of doping agent aggregate.}
	\label{fig:Circulation}
\end{figure}

Past studies, see for example~\cite{pandey2019high}, suggest two distinct mechanisms lead to mass loss from the main droplet during Phase 1 (when atomization is not present). These are gasification of liquid fuel at the droplet surface (which is present for both pure and doped ethanol) and permeation of the liquid fuel through a porous shell of nanoparticles formed near the droplet surface. Gan et al. \cite{gan2011combustion} suggest that the fluid motion, e.g. the droplet surface regression, the droplet expansion as a result of bubble formation, and the internal flow circulation may lead to the formation of the porous shell. Although our measurements do not allow for imaging the shell formation, evidence of flow circulation inside the droplet is present in our study. Figure~\ref{fig:Circulation} shows a series of shadowgraphy images that underline the movement of aggregated particles inside the droplet due to a possible internal flow circulation. The results in the figure pertain to Er0.01 test condition. Figures \ref{fig:Circulation}(a--d) correspond to $t = 2200$, 2275, 2350 and 2425~ms, respectively. As can be seen, a nanomaterial aggregate highlighted by the red dashed circle in the figure undergoes a clockwise motion, possibly due to an internal circulatory flow inside the droplet. During Phase 2, in addition to gasification and possible permeation of fuel through the nanomaterial porous shell, mass loss due to atomization is also present. Thus, the larger absolute value of the slope of $D^2$ versus time was observed compared to that of Phase 1. \par

\begin{figure}[!t]
	\centering
	\includegraphics[width=1\textwidth]{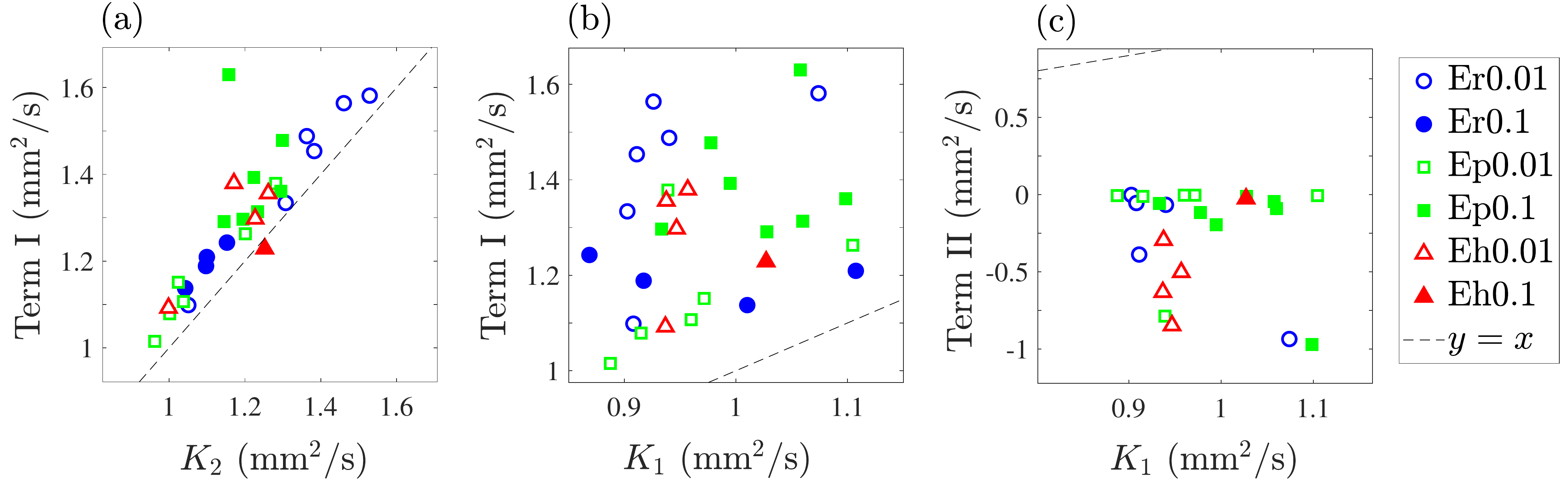} 
	\caption{Variation of (a) Term I versus $K_2$, (b) Term I versus $K_1$, and (c) Term II versus $K_1$.}
	\label{fig:Term1andK2}
\end{figure}

It is of interest to quantify the effect of atomization on the burning rate of the doped ethanol droplets. Assuming that the total mass loss during the droplet lifetime is due to baby droplets ejection and evaporation mechanisms, conservation of mass suggests

\begin{equation}
	\label{eq:massconservation1}
	\Delta m_{\mathrm{T}} = \Delta m_{\mathrm{EJ}} + \Delta m_{\mathrm{EV}},
\end{equation}
where $\Delta m_{\mathrm{T}}$, $\Delta m_{\mathrm{EJ}}$, and $\Delta m_{\mathrm{EV}}$ are the total mass loss, the mass loss due to the baby droplet ejections, as well as the mass loss due to evaporation. It is important to highlight that $\Delta m_{\mathrm{EJ}}$ only pertains to Phase 2; however, $\Delta m_{\mathrm{EV}}$ is present during the entire droplet lifetime. The total mass loss depends on the droplet density and the main droplet initial equivalent diameter, given by $\Delta m_\mathrm{T}=-\rho_{\mathrm{l}} \pi D_{0}^{3}/6$, with $\rho_{\mathrm{l}}$ being pure ethanol density. In fact, results of~\cite{keblinski2005nanofluids,ooi2017improving,ghamari2016experimental,basu2016combustion,sabourin2010exploring,kim2007surface,chen2007rheological,tanvir2012surface} suggest that the droplet density does not change substantially for the loading concentrations studied in the present investigation. The first term on the right-hand-side (RHS) of Eq.~(\ref{eq:massconservation1}) is given by 

\begin{equation}
	\label{eq:atomization}
	\Delta m_{\mathrm{EJ}}=-\sum_{i=1}^{N}\dfrac{\rho_\mathrm{l} \pi}{6} d_{0}^3(i),
\end{equation}
where $d_0(i)$ is the initial diameter of the $i^\mathrm{th}$ detected baby droplet, and $N$ is the total number of the detected baby droplets. In Eq.~(\ref{eq:atomization}), it is inherently assumed that the concentration of the doping agent inside the baby droplets remains small (similar to the main droplet). Thus, the density of the baby droplets equals the ethanol density. Assuming that the droplet burning rates during the heat-up ($0 \leq t \leq 0.05\tau$) and flame extinction ($0.95\tau \leq t \leq \tau$) periods equate to those in Phases 1 and 2, respectively, $\Delta m_{\mathrm{EV}}$ is given by:

\begin{equation}
	\label{eq:massconservation2}
	\Delta m_{\mathrm{EV}} \approx \int_{0}^{\tau}{\dot{m}_{\mathrm{EV}}(t)\mathrm{d}t},
\end{equation}
where $\dot{m}_{\mathrm{EV}}(t)$ is the time rate of change of the main droplet mass due to evaporation. Assuming that the liquid density remains constant during the droplet evaporation, it can be shown that
\begin{equation}
	\label{eq:massconservation3}
	\dot{m}_{\mathrm{EV}}(t)=\frac{\rho_{\mathrm{l}} \pi}{4} D'(t) \left[\frac{\mathrm{d}D'^2(t)}{\mathrm{d}t}\right],
\end{equation}
where $D'(t)$ is the main droplet equivalent diameter provided that the baby droplet ejections are absent. Only during Phase 1, $D'(t) = D(t)$ since the baby droplet ejections do not occur during this phase; however, $D'(t) \neq D(t)$ during Phase 2. It is important to highlight that $D'(t)$ pertaining to Phase 2 cannot be measured experimentally, since the atomization is present during Phase 2. Following the above discussion, Eqs.~(\ref{eq:massconservation2})~and~(\ref{eq:massconservation3}) were combined and the integral in Eq.~(\ref{eq:massconservation2}) was expanded and is given by

\begin{equation}
	\label{eq:massconservation4}
	\Delta m_{\mathrm{EV}} \approx \frac{\rho_{\mathrm{l}} \pi}{4}\left[\int_{0}^{\tau_{\mathrm{a}}}{D(t) \underbrace{\left(\frac{\mathrm{d}D^2(t)}{\mathrm{d}t}\right)}_{-K_1}\mathrm{d}t}+ \int_{\tau_{\mathrm{a}}}^{\tau}{D'(t) \underbrace{\left(\frac{\mathrm{d}D'^2(t)}{\mathrm{d}t}\right)}_{-K'_2}\mathrm{d}t}\right],
\end{equation}
with $\tau_{\mathrm{a}}$ being the time instant at which the first baby droplet appears, referred to as the atomization time delay, see Fig.~\ref{fig:SGcurve}(b). During Phase 1, baby droplet ejections are absent and $\mathrm{d}D^2(t)/\mathrm{d}t = -K_1$. In Eq.~(\ref{eq:massconservation4}), $\mathrm{d}D'^2(t)/\mathrm{d}t$ pertains to only the droplet evaporation, does not equate to $-K_2$ shown in Fig.~\ref{fig:SGcurve}, and is referred to as $-K'_2$. In other words, the time rate of change of $D'^2$ is the contribution of evaporation to droplet burning rate only due to evaporation during Phase 2. Using the above discussions, it is straightforward to show that $D(t)$ (pertaining to Phase 1) and $D'(t)$ (pertaining to Phase 2) can be obtained from

\begin{subequations}\label{eq:massconservation5}
	\begin{equation}\label{eq:massconservation5a}
		\begin{array}{ll}
			D(t) \approx \sqrt{D_0^2-K_{\mathrm{1}}t},  & 0 \le t < \tau_{\mathrm{a}},
		\end{array}
	\end{equation}
	\begin{equation}\label{eq:massconservation5b}
		\begin{array}{ll}
			D'(t) \approx \sqrt{D_{\mathrm{a}}^2-K'_{\mathrm{2}}(t-\tau_{\mathrm{a}})}, & \tau_{\mathrm{a}} \le t \le \tau.
		\end{array}
	\end{equation}
\end{subequations}
Combining Eqs.~(\ref{eq:massconservation1}),~(\ref{eq:atomization}),~(\ref{eq:massconservation4})~(\ref{eq:massconservation5a}),~and~(\ref{eq:massconservation5b}), it can be shown that $K'_{\mathrm{2}}$ is given by

\begin{equation}
	\label{eq:massconservation6}
	K'_{\mathrm{2}} \approx \underbrace{\left(\frac{D^{\mathrm{2}}_{\mathrm{a}}}{\tau-\tau_{\mathrm{a}}}\right)}_{\mathrm{Term~I}} + \underbrace{\left(\frac{-\sum_{i=1}^{N}d^{\mathrm{2}}(i)}{\tau-\tau_{\mathrm{a}}}\right)}_{\mathrm{Term~II}}.
\end{equation}
In Eq.~(\ref{eq:massconservation6}), Term I is similar to $K_2$, except this term pertains to $\tau_\mathrm{a} \lesssim t \lesssim \tau$, but $K_2$ is obtained using a linear fit to the droplet equivalent diameter squared for $\tau_\mathrm{a} \lesssim t \lesssim 0.95 \tau$. Variation of Term I versus $K_2$ is shown in Fig.~\ref{fig:Term1andK2}(a). In the figure, open and solid symbols pertain to GO concentration of 0.01 and 0.1~\%, respectively. As can be seen, the results nearly follow the line of $y = x$, confirming the assumption made earlier that the droplet burning rate during Phase 2 and during $0.95\tau \lesssim t \lesssim \tau$ are similar. Variations of Term I and Term II versus $K_1$ are presented in Figs.~\ref{fig:Term1andK2}(b) and (c), respectively. The results in Fig.~\ref{fig:Term1andK2}(b) suggest that Term I is always larger than $K_1$, which is expected and is due to the combined effects of atomization and evaporation. Term II, however, is negative and can become significantly small (about -1~$\mathrm{mm^2/s}$), see Fig.~\ref{fig:Term1andK2}(c).

\begin{figure}[!t]
	\centering
	\includegraphics[width=0.5\textwidth]{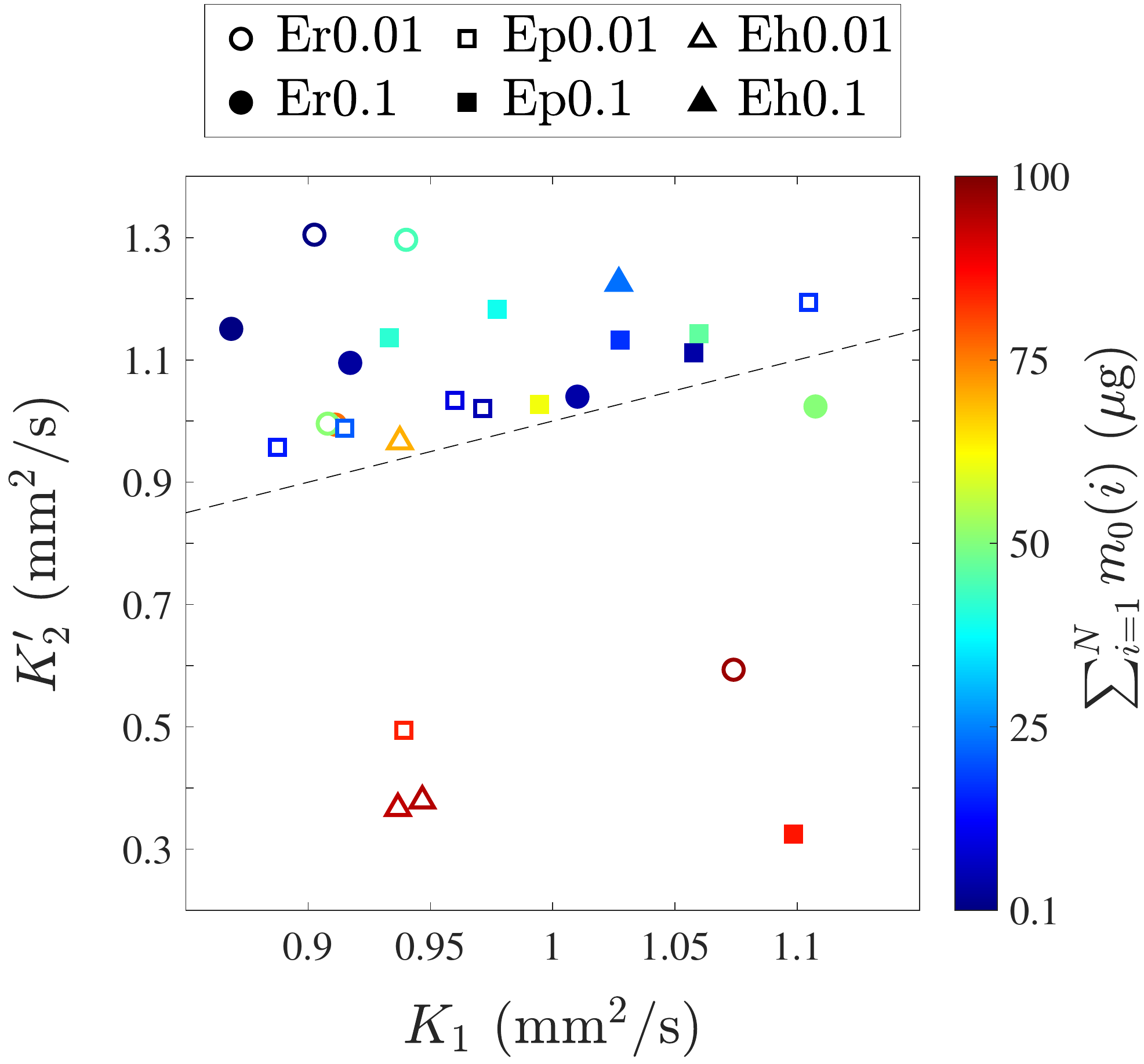} 
	\caption{$K'_2$ versus $K_1$ for the cases that atomization is present. The figure is color-coded based on the total mass of the ejected baby droplets. The dashed line is $y = x$.}
	\label{fig:link}
\end{figure}

It is of interest to investigate the effect of atomization on the rate of evaporation during Phase 2. Thus, the variation of $K'_2$ versus $K_1$ is presented in Fig.~\ref{fig:link}, and the results are color-coded based on the total mass loss due to baby droplets ejections. As can be seen, for a relatively small amount of baby droplet ejections (shown with dark blue symbols), the rate of evaporation during Phase 2 is close to that of Phase 1. However, as the total mass of the ejected baby droplets increases, $K_1$ decreases to values as small as 0.3~$\mathrm{mm^2/s}$, see the red data symbols in Fig.~\ref{fig:link}. This means that, for significant amounts of the baby droplet ejections, the evaporation rate is suppressed. Nonetheless, the combined effect of mass loss due to evaporation and atomization is such that $K_2 (\mathrm{or~Term~I}) \gtrsim K_1$. 
\par

\begin{figure} [!t]
\centering
\includegraphics[width=1\textwidth]{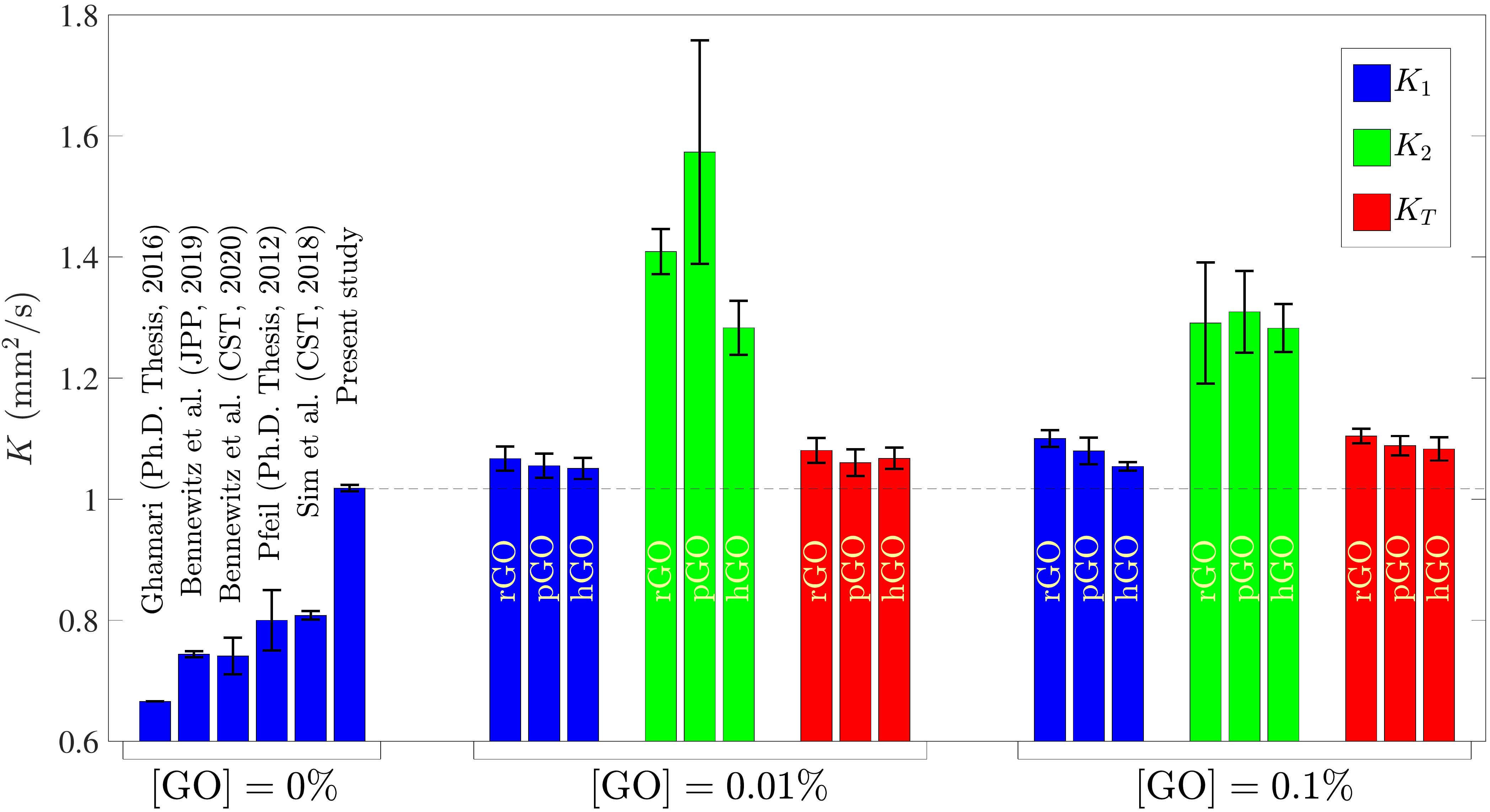} 
\caption{Effect of graphene oxide type and concentration on the burning rate of ethanol droplets.}
\label{fig:K}
\end{figure}

The averaged values of $K_1$ and $K_2$ for all tested conditions are presented in Fig.~\ref{fig:K} using blue and green bars, respectively. The values of pure ethanol burning rate were extracted from the studies of~\cite{ghamari2017combustion,bennewitz2019systematic,bennewitz2020combustion,sim2018effects,pfeil2012changes} and are presented in the figure by the blue bars as well. In order to facilitate comparisons, as well as to assess the overall burning rate of the doped droplets, a line was fit to the equivalent diameter squared regression curves, see the red dashed lines in Fig.~\ref{fig:SGcurve}(b). The absolute value of the slope of the line is referred to as $K_\mathrm{T}$, and the values are presented in Fig.~\ref{fig:K} using red bars. The results in the figure show that the reported burning rate of ethanol in the present study is about 35\% larger than the average value of the results from the literature. This is due to the difference between the droplet suspension mechanism used in the present study and those used in \cite{ghamari2017combustion,bennewitz2019systematic,bennewitz2020combustion,sim2018effects,pfeil2012changes}. It is speculated that the relatively large diameter of the supporting mechanism used in the present study enhances heat transfer from the supporting mechanism to the droplet. In fact, past investigations~\cite{chauveau2019analysis,yang2002experimental,liu2015effect,chauveau2008experimental} reported that increasing the diameter of the supporting mechanism increases the burning rate.
\par

Although the droplet supporting mechanism can influence the burning rate, since the supporting mechanism geometry is fixed in this study, comparisons can be made between the conditions tested here. The results show that increasing the loading concentration and reducing the oxygen content of graphene oxide both increase $K_1$. That is doping pure ethanol with 0.1\% of reduced graphene oxide leads to the maximum enhancement of about 8\% in $K_1$. It can be inferred from the results presented in~\cite{taylor2013small,keblinski2005nanofluids,mahian2019recent,chen2020effect} that increasing the concentration of the doping agent as well as reducing the oxygen content of the graphene oxide increase the thermal conductivity of the fuel. It is speculated that the enhancement of $K_1$ with addition of graphene oxide is due to a potential increase of the thermal conductivity, which is similar to the conclusions reported in~\cite{sabourin2009functionalized,singh2019effect,singh2020effect,aboalhamayie2019evaporation}.

\begin{figure} [!t]
	\centering
	\includegraphics[width=0.5\textwidth]{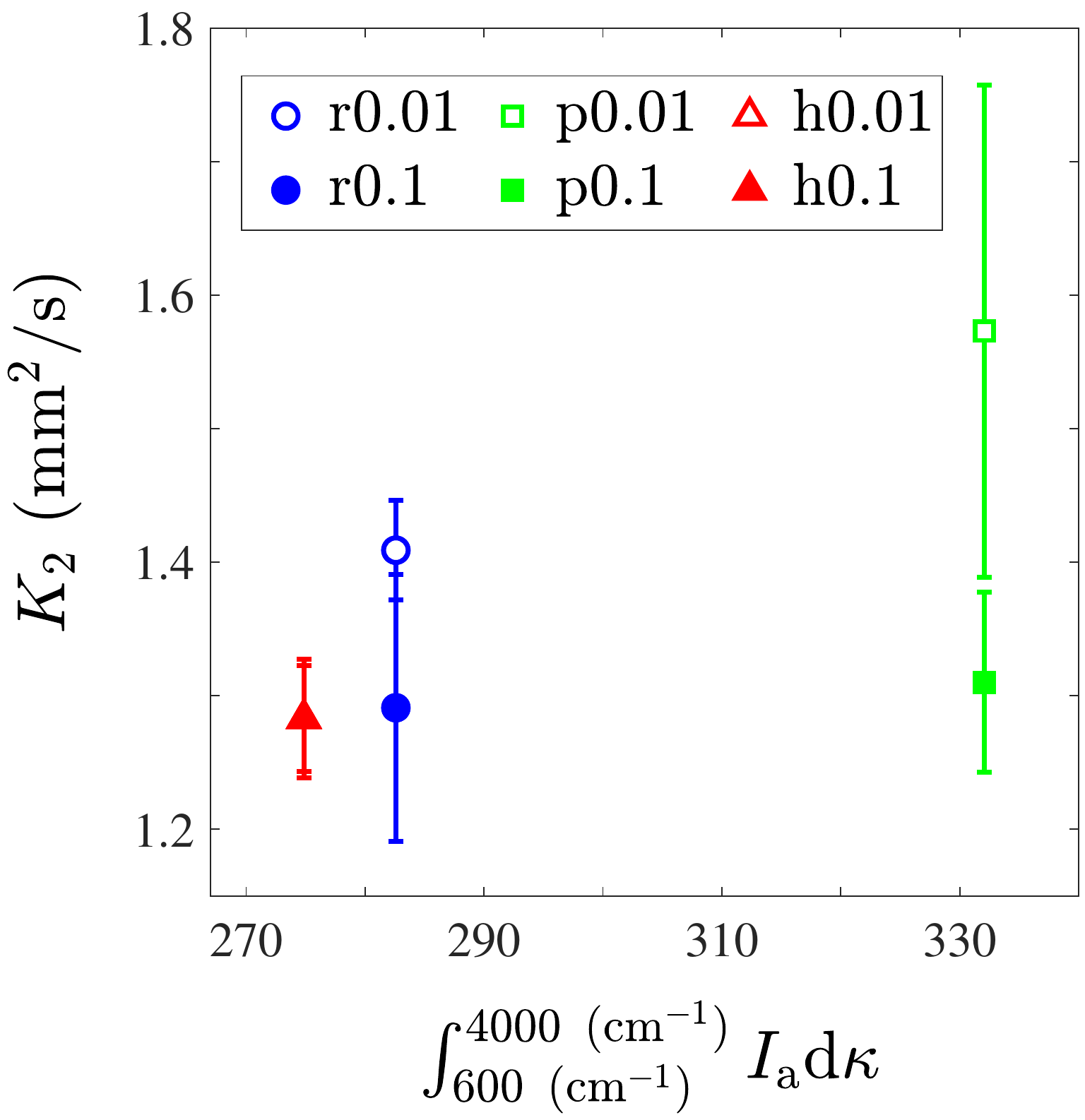} 
	\caption{Variation of $K_2$ versus the area under the FT-IR curves (see Fig.~\ref{fig:FT-IR}) for ethanol doped with rGO, pGO, and hGO.}
	\label{fig:absorption}
\end{figure}

The results presented in Fig.~\ref{fig:K} show that the values of $K_2$ are larger than $K_1$ for all tested conditions. As can be seen, for the loading concentration of 0.01\%, the largest value of $K_2$ pertains to pGO, followed by hGO and rGO. As discussed in~\cite{miglani2015coupled}, adding the doping agent increases the number of sites prone to absorption of radiation heat, which leads to the formation of bubbles, and as a result the atomization. As discussed in \cite{gaydon2012spectroscopy}, most of the flame thermal radiation may correspond to the infrared wavelength ranging from about 1 to 24~$\mathrm{\mu}$m, which relates to the wave-number varying from 400 to 10000~$\mathrm{cm^{-1}}$. This is mostly due to radiation from $\mathrm{CO_2}$ and $\mathrm{H_2O}$ \cite{gaydon2012spectroscopy}. The FT-IR results related to the radiation absorption of the tested nanomaterials allowed for characterizing the emission absorption versus the wave-number ranging from 600 to 4000~$\mathrm{cm^{-1}}$, see Fig.~\ref{fig:FT-IR}. Variation of $K_2$ versus the area under the absorption curves in Fig.~\ref{fig:FT-IR} is presented in Fig.~\ref{fig:absorption}. The blue, green, and red data symbols in the figure correspond to ethanol droplets doped with rGO, pGO, and hGO, respectively. In the figure, open and solid symbols present 0.01 and 0.1\% by weight concentrations of the doping agent, respectively. As can be seen, increasing the area under the absorption curves is accompanied by an increase in the values of $K_2$ for the smallest tested concentration; however, this increase is not significant for the largest tested concentration. The results in Fig.~\ref{fig:absorption} suggest that the burning rate in Phase 2 can be significantly increased for ethanol doped with GO by increasing this material radiation absorption. However, this improvement depends on the doping concentration. In fact, Basu and Miglani~\cite{basu2016combustion} argued that the increase in the loading concentration might lead to the formation of a stronger aggregation shell which could resist the pressure build-up (as a result of bubble formation) and could potentially suppress the ejection events. This is consistent with the results shown in Figs.~\ref{fig:K}~and~\ref{fig:absorption} for the loading concentration of 0.1\% compared to 0.01\%. 

The results show that adding the carbonaceous doping material increases the overall burning rate of ethanol. This is of significant importance for the industrial application of such nanofuels. Specifically, the results in Fig.~\ref{fig:K} suggest that, for the loading concentration of 0.01\%, the addition of rGO, pGO, and hGO to ethanol leads to an enhancement of the overall burning rate by 6.2, 4.8, and 4.1\%, respectively. Similarly, for the loading concentration of 0.1\%, adding rGO, pGO, and hGO results in an enhancement of 8.4, 6.9, and 6.4\% of the overall burning rate, respectively. Thus, it is concluded that the maximum enhancement of about 8\% can be achieved by adding 0.1\% of rGO to ethanol droplets.
\par

\section{Concluding remarks}
\label{conclusion}

Ignition delay and burning rate of graphene oxide doped ethanol droplets were experimentally investigated using separate high-speed $\mathrm{OH^*}$ chemiluminescence and high-speed shadowgraphy imaging, respectively. Reduced, partially, and highly oxidized graphene oxides were synthesized and two loading concentrations (by weight) of 0.01 and 0.1\% were tested in the experiments.

The highly oxidized graphene (hGO) was thermally treated for two different time periods, producing partially oxidized (pGO) and reduced graphene oxide (rGO) samples. Raman spectroscopy, high- and coarse-resolution Scanning Electron Microscopy, Fourier-Transform Infrared spectroscopy, as well as thermogravimetric analysis were performed to characterize the synthesized graphene oxide nanomaterials. The Raman spectroscopy results show that the synthesized nanomaterials are, indeed, graphene oxide. The results of high-resolution scanning electron microscopy illustrated the layered structure of graphene oxide and showed that the thickness the stack of 2D layers is smaller than about 100 nm. Coarse-resolution scanning electron microscopy was performed to estimate the probability density function of the characteristic length of the tested graphene oxide nanomaterials. It was shown that the probability density function of the characteristic length is similar for all three types of the synthesized nanomaterials, with the most probable characteristic length pertaining to about 2~$\mathrm{\mu}$m. Fourier-Transform Infrared spectroscopy of the three samples featured peaks at similar wave-numbers and suggested the existence of matching oxygen-containing functional groups. The combination of the results pertaining to the infrared spectroscopy and the thermogravimetric analysis suggested that rGO, pGO, and hGO, indeed, correspond to graphene oxide with least, moderate, and maximum amounts of oxidation, respectively. \par

The estimated ignition delay for pure ethanol in the present study was $158.5 \pm 6.8$~ms. However, this parameter is reported to vary between 71 to 274~ms in the literature. The normalization of the ignition delay with the droplet surface area or with the droplet volume did not allow for the collapse of the pure ethanol ignition delay reported in the present study and those reported in the literature. It was concluded that, in addition to the droplet size, ignition mechanism, the droplet supporting mechanism, and the droplet size also influence the estimated ignition delay. For ethanol droplets doped with rGO and pGO, the ignition delay increases almost linearly with increasing the loading concentration from 0.01 to 0.1\%. However, ethanol droplets doped with hGO feature a maximum ignition delay at the doping concentration of 0.01\%. For all tested conditions, a maximum increase of about 21\% was observed in the ignition delay of ethanol doped with pGO, for the loading concentration of 0.1\%. \par

The results suggest that the atomization phenomenon could occur for all tested doped fuels. However, this phenomenon did not take place for pure ethanol droplets. The results show that for all experiments that feature the atomization phenomenon, the baby droplet ejections only take place in the second half of the droplet lifetime. It is shown that the most probable values of the baby droplets initial velocity projected in the $x-y$ plane, the baby droplets initial diameter, and the length of the baby droplets trajectory projected in the $x-y$ plane are about 0.5~m/s, 50~$\mathrm{\mu}$m, and 1~mm, and are independent of the nanomaterial type and the tested loading concentration. The most probable value of the baby droplet initial velocity and length of trajectory are comparable to the ethanol and air mixture flame speed at the stoichiometric condition and the size of the main droplet. The joint probability density function of the above-mentioned parameters against one another was calculated. It was shown that most of the baby droplets initially feature relatively small diameter and travel about 1~mm with an initial projected velocity of about 0.5~m/s in the $x-y$ plane. This finding highlights that most of the baby droplets feature a relatively short lifetime and may potentially burn before escaping the flame. \par

It was shown that the absolute value of the slope of the droplet equivalent diameter squared versus time, which is the burning rate in the corresponding period, increases to a large and fixed value immediately after the first baby droplet is detected. The absolute value of the mentioned slope in the period that atomization does not occur and in the period that atomization takes place were referred to as $K_1$ and $K_2$, respectively. The burning rate analysis showed that increasing the loading concentration and reducing the oxidation level of the doping agents increase $K_1$. This enhancement is speculated to be related to the improvement of the thermal conductivity of the fuel due to the addition of the nanomaterials. Variation of $K_2$ among the fuels doped with different graphene oxide samples suggested that pGO features the largest values of $K_2$, followed by rGO and hGO, respectively. The Fourier-Transform infrared spectroscopy results suggested that this trend is linked to the amount of thermal radiation absorption of the doping agents. Specifically, it was shown that for a loading concentration of 0.01\%, $K_2$ increases almost linearly with increasing the area under the curve of the Fourier-Transform infrared spectroscopy. Such an increase for $K_2$ also exists for loading concentration of 0.1\%. However, the amount of increase of $K_2$ was smaller compared to that for the loading concentration of 0.01\%. A mass-balance framework was used to calculate the effect of atomization on the droplet mass loss. It was obtained that, in the presence of relatively intense atomization (which pertains to a relatively large total mass of the baby droplets), the evaporation mass loss is suppressed. The results suggest that the addition of graphene oxide into ethanol and increasing the loading concentration from 0.01 to 0.1\% improves the overall burning rate. A maximum improvement of 8.4\% was observed as a result of rGO addition for the loading concentration of 0.1\%.

\section*{Acknowledgements}
The authors are grateful for financial support from the Natural Sciences and Engineering Research Council (NSERC) Canada through the Collaborative Research and Development grant (CRDPJ 536828-18). Financial support as well as raw graphite material provided by ZEN Graphene Solutions are gratefully acknowledged. The authors would also like to thank Dr. Ehsan Ebrahimnia-Bajestan, Dr. Gethin Owen, Dr. Sudip Shrestha, Dr. Seyyed Arash Haddadi, and Negin Jalili for their assistance with the nanomaterials preparation, SEM imaging, Raman spectroscopy, FT-IR, and TGA experiments.

\section*{Appendix A: Effect of igniter retraction motion on the flame}
\label{ApA}
\begin{figure}
	\centering
	\includegraphics[width=1\textwidth]{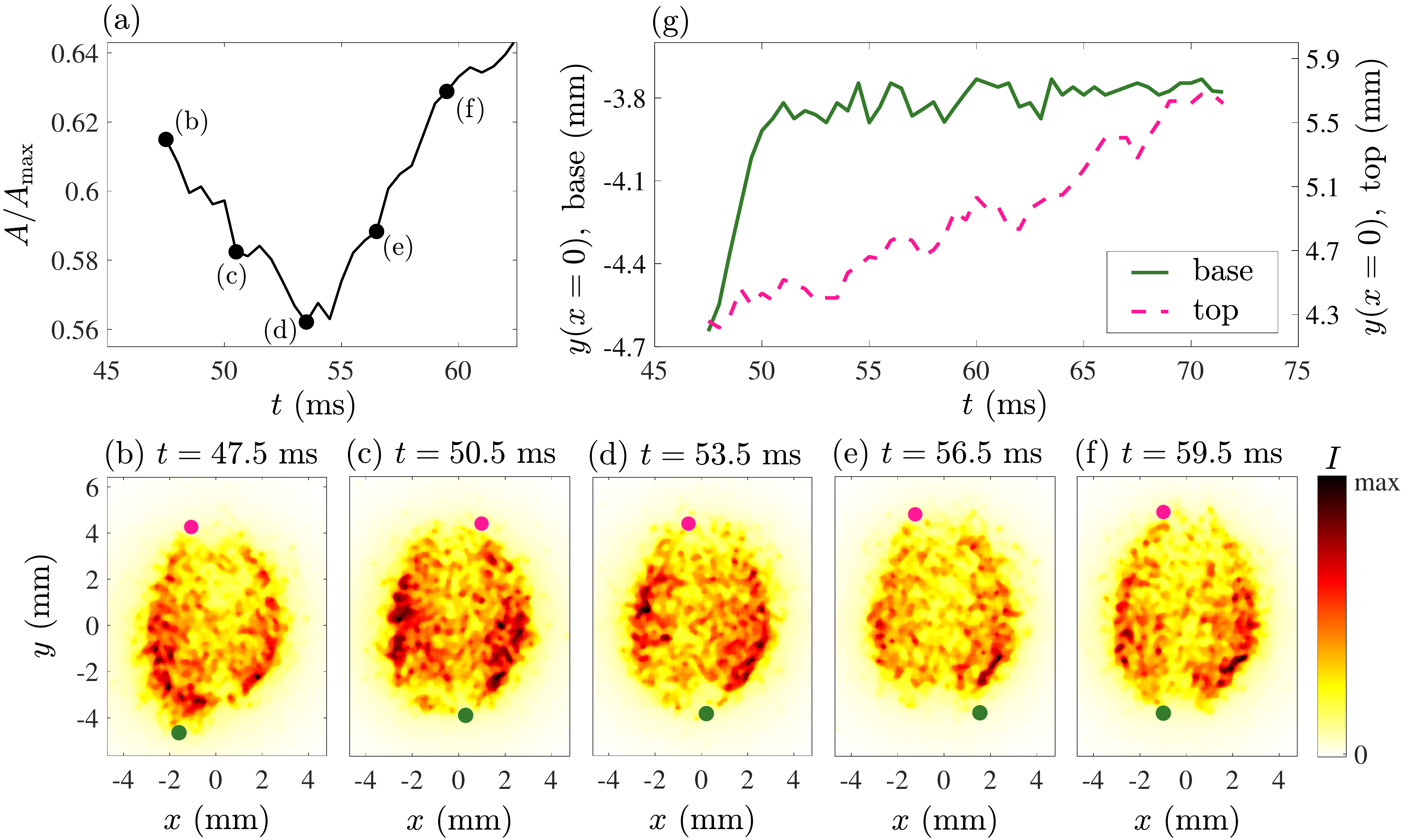} 
	\caption{(a) Normalized flame area versus time, (b--f) correspond to the $\mathrm{OH^*}$ images at $t = 47.5$, 50.5, 53.5, 56.5, and 59.5~ms, respectively. (g) presents the location of the flame base (green solid curve) and the top of the flame (dashed pink curve) versus time. The results pertain to the pure ethanol test condition.}
	\label{fig:PA}
\end{figure}

Variation of $A/A_\mathrm{max}$ pertaining to the test condition of pure ethanol is shown Fig.~\ref{fig:PA}(a). The flame chemiluminescence pertaining to $t = 47.5$, 50.5, 53.5, 56.5, and 59.5~ms are presented in Figs.~\ref{fig:PA}(b--d), respectively. During the time period shown in Fig.~\ref{fig:PA}(a), $A/A_\mathrm{max}$ is influenced by the retracting motion of the igniter. In order to demonstrate this, the global maximum of flame chemiluminescence during $47.5~\mathrm{ms} \lesssim t \lesssim 59.5~\mathrm{ms}$ was identified, and the flame chemiluminescence contour corresponding to 10\% of its maximum was obtained. Then, the vertical locations of the contour at $x = 0$ were identified and used to highlight the flame base and top, which are shown by the green and pink data points in Figs.~\ref{fig:PA}(b--f), respectively. These locations are also shown by the green solid and pink dashed curves in Fig.~\ref{fig:PA}(g). As can be seen, for about 5~ms, the flame base moves upward faster than the flame top, which decreases the flame size, hence the decrease of $A/A_\mathrm{max}$ shown in Fig.~\ref{fig:PA}(a) during $47.5 \lesssim t \lesssim 53.5$~ms. Such decreasing trend is induced by the igniter retracting motion, pertains to a relatively short time period (5~ms), is independent of the tested condition, and is relatively negligible compared to the ignition delay. Thus, it is concluded that the igniter retracting motion does not substantially influence the flame burning dynamics.

\section*{Appendix B: Spatial resolution of the shadowgraphy measurements}
\begin{figure}
	\centering
	\includegraphics[width=0.5\textwidth]{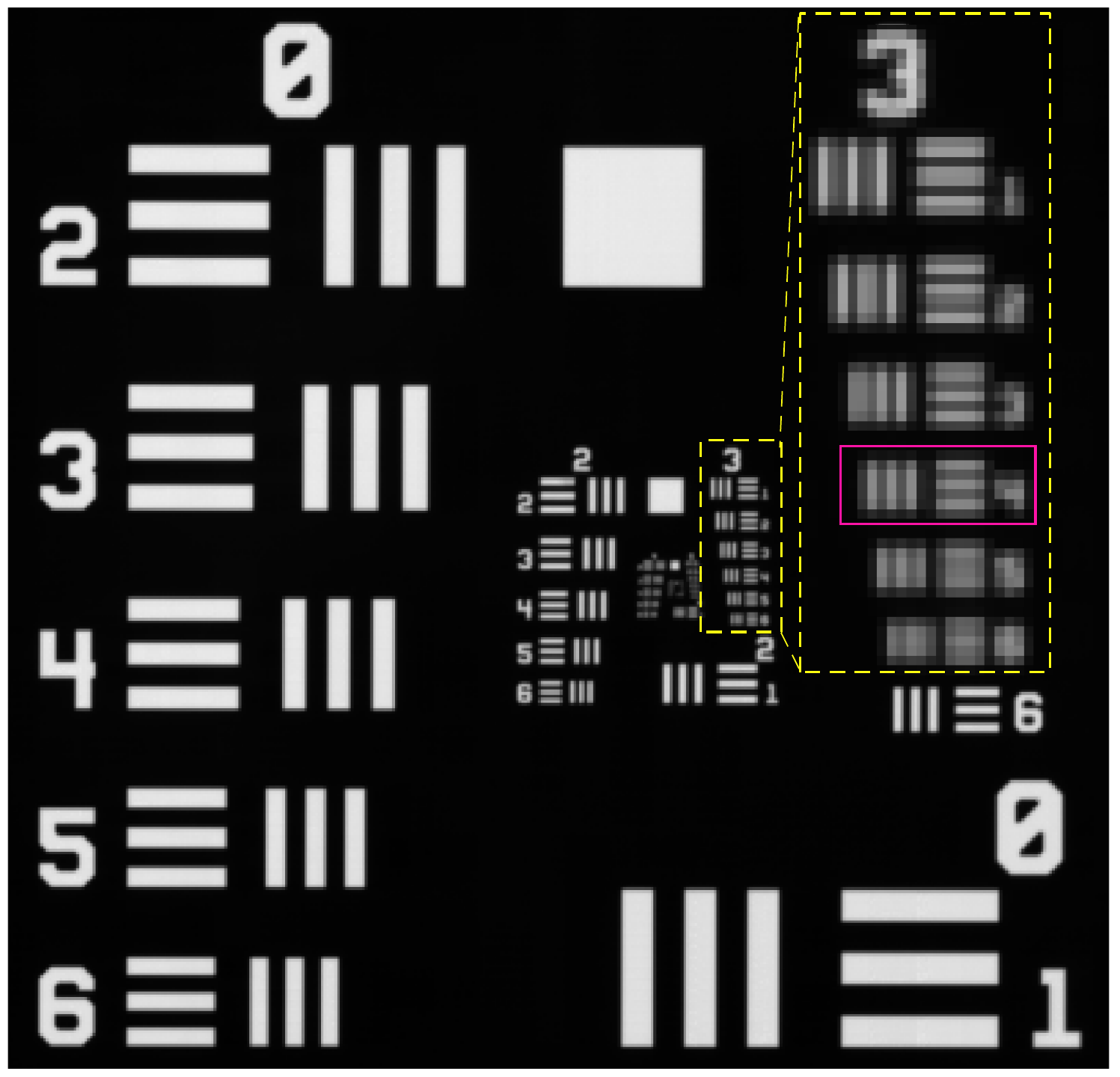} 
	\caption{Image of the USAF 1951 target acquired with settings identical to those used in the shadowgraphy experiments.}
	\label{fig:USAF}
\end{figure}

Spatial resolution of the shadowgraphy measurements is estimated using the USAF 1951 resolution target, similar to \cite{papageorge2014recent}. An image of the target is shown in Fig.~\ref{fig:USAF}. Identical camera settings to those used for the shadowgraphy measurements were used for acquiring the image. In the figure, the yellow-dashed rectangle highlights the lines whose thickness is close to the pixel resolution, i.e. the side length of the field-of-view divided by the corresponding number of pixels ($39~\mathrm{mm}/1024~\mathrm{pixels} = 38~\mu~\mathrm{mm/pixel}$). The pink dashed box highlights the white rectangles whose width is almost equal to the shadowgraphy images pixel resolution. \par

\begin{figure}[!t]
	\centering
	\includegraphics[width=0.9\textwidth]{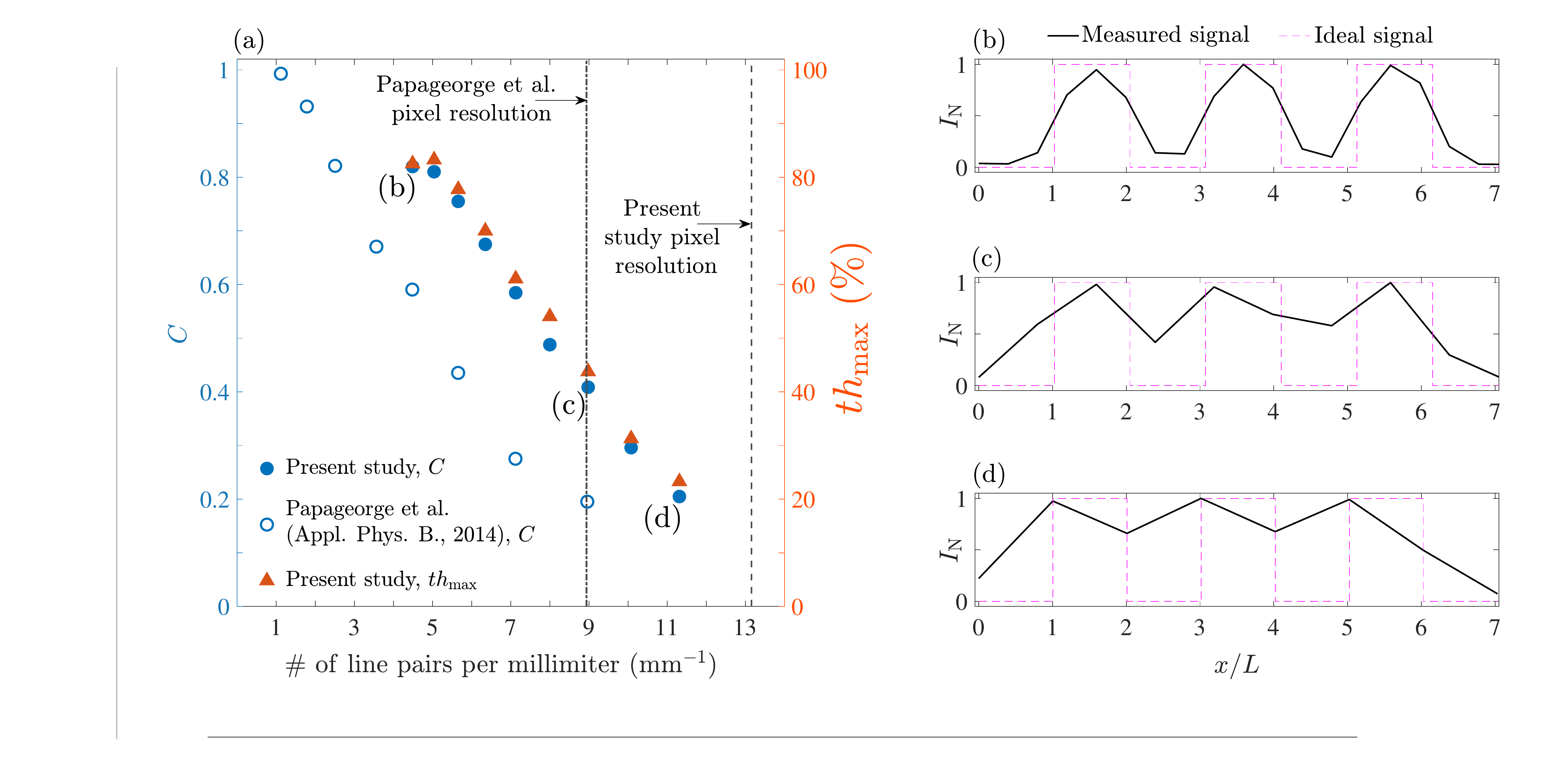} 
	\caption{(a) Variation of contrast ($C$) versus the number of line pairs per millimeter for the present study and that of \cite{papageorge2014recent}. The vertical axis on the right hand side is the maximum threshold required to distinguish the white rectangles. (b--d) are the variations of $I_\mathrm{N}$ across the withe rectangles versus $x/L$ for $L = 111.4$, 55.4, and 44.2~$\mu \mathrm{m}$, respectively.}
	\label{fig:LSF}
\end{figure}

In order to estimate the imaging resolution, first, a parameter referred to as contrast ($C$) was estimated using \cite{papageorge2014recent}
 
\begin{equation}
	\label{eq:contrast}
	C=\dfrac{I_\mathrm{max}-I_\mathrm{min}}{I_\mathrm{max}+I_\mathrm{min}},
\end{equation} 
where $I_\mathrm{max}$ and $I_\mathrm{min}$ are the maximum and minimum light intensity ($I$) acquired across each group of three white rectangles shown in Fig.~\ref{fig:USAF}. Variations of the contrast versus the number of line pairs (combination of one white and black rectangles) in one millimeter calculated in the present study and extracted from \cite{papageorge2014recent} are shown by the solid and open blue circular data symbols, respectively, in Fig.~\ref{fig:LSF}(a). As expected, increasing the number of line pairs per one millimeter (i.e. selecting smaller thickness lines) decreases the ability of the camera to distinguish the lines. The extent of the horizontal axis is the inverse of the twice of the field-of-view resolution. Papageorge et al.~\cite{papageorge2014recent} used $C = 0.2$ to estimate their imaging resolution. The value of the imaging resolution at $C = 0.2$ is approximately 44~$\mu$m in the present study.

Figures~\ref{fig:LSF}(b--d) show the variations of light intensity normalized by the corresponding maximum intensity ($I_\mathrm{N} = I/I_\mathrm{max}$) versus $x$ normalized by the corresponding white rectangle width ($L$). Results in (b), (c), and (d) correspond to $L = 111.4$, 55.4, and 44.2~$\mu \mathrm{m}$, respectively. In Figs.~\ref{fig:LSF}(b--d), the purple-dashed lines are used to mark perfect signals with least modulations. The results in Fig.~\ref{fig:LSF}(d) show that for the line pairs with $L = 44.2$, three distinct peaks related to three white rectangles are detected. This means that, for the resolution of about 44~$\mu \mathrm{m}$ (which is larger than but close to the pixel resolution), the true signal can be detected. 

The results presented in Fig.~\ref{fig:LSF}(a) suggest that baby droplets as small as nearly one pixel may be detected in our measurements. However, such ability is substantially influenced by the post-processing of the collected images. Specifically, the threshold value used for binarization of the images as well as the size of the filter influence the ability to detect the baby droplets. The technique used here for detection of the baby droplets is similar to the method presented in~\cite{bradley2007adaptive}. First, an algorithm is developed in MATLAB to detect candidate baby droplets. Then, the center of each candidate baby droplet is obtained, a $w\times w~\mathrm{pixel^2}$ window is considered around the baby droplet center, and the mean intensity inside the window ($M_w$) is calculated. Finally, pixels whose intensities are larger than $M_w(1-th)$ are disregarded, with $th$ being the selected threshold. Using a similar approach for the white rectangles, the maximum threshold ($th_\mathrm{max}$) that allows for detecting the white rectangles was obtained and presented against the number of line pairs per millimeter in Fig.~\ref{fig:LSF}(a) using the orange triangular data points. As can be seen, at the imaging resolution of 44~$\mu \mathrm{m}$, the maximum threshold that allows for detecting the white rectangles is about 23\%. On one hand, values of $th$ larger than 23\% removes the white rectangles from the post-processed images and are not recommended to be used. On the other hand, too small threshold values does not allow to distinguish between the background noise and the rectangles. Inspecting the sequence of images captured from the shadowgraphy technique, it was obtained that $th \lesssim 15~\mathrm{\%}$ can lead to detection of noise as baby droplet. Thus, the selected threshold should vary between 15\% and 23\% in the present study.

\begin{figure}[!h]
	\centering
	\includegraphics[width=0.5\textwidth]{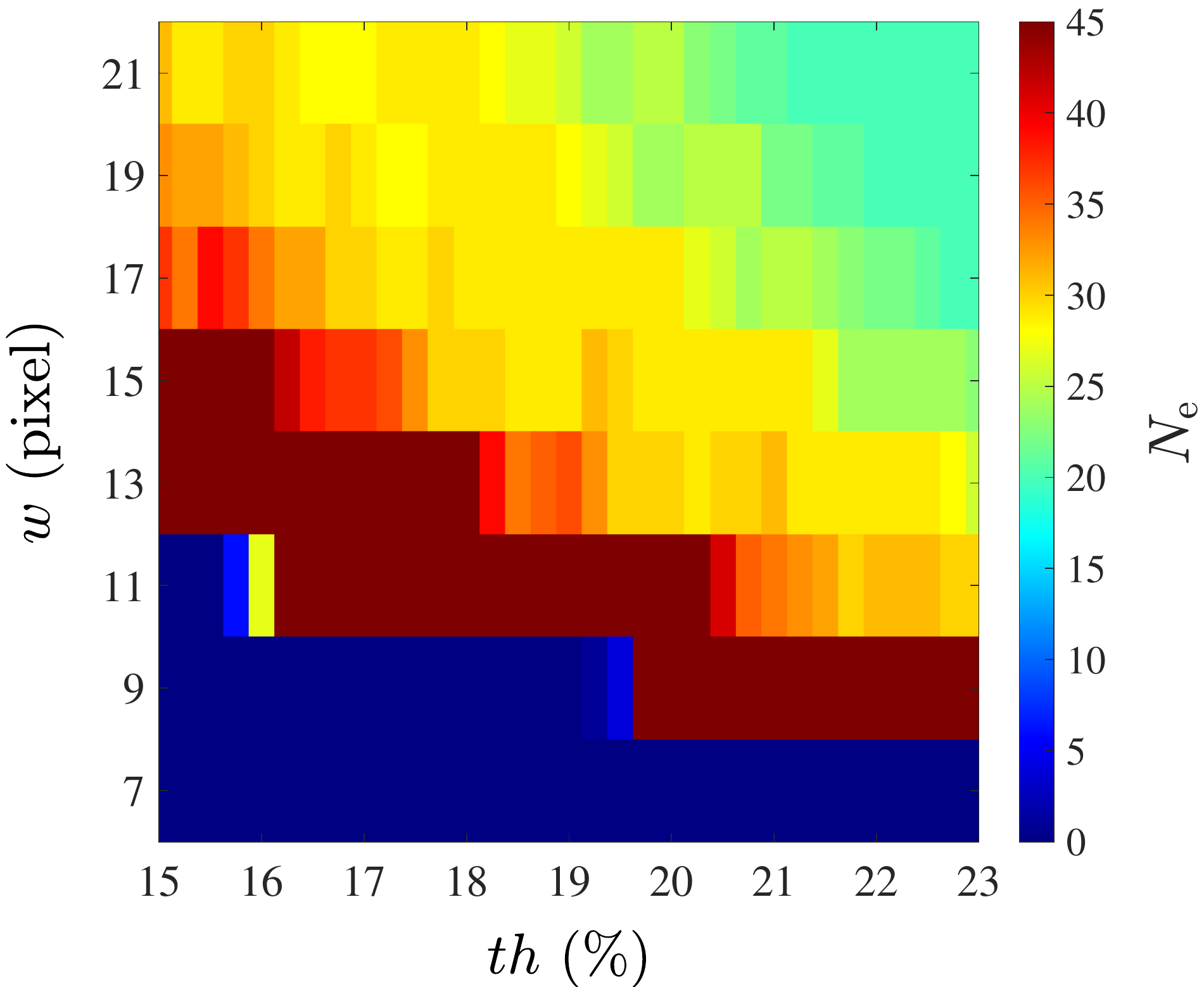} 
	\caption{Effects of the threshold and window size on the number of ejected baby droplets. The results pertain to the test condition of Eh0.01.}
	\label{fig:thw}
\end{figure}

For the above range of threshold (i.e. $15\% \lesssim th \lesssim 23\%$), the effects of both $th$ and $w$ on the number of ejected baby droplets ($N_\mathrm{e}$) were assessed. Exemplary results related to the test condition of Eh0.01 are presented in Fig.~\ref{fig:Ejectionevent}. The window size ($w\times w$) was varied from $7\times 7$ to $21\times 21~\mathrm{(pixel^2)}$. After careful examination of the raw images, it was obtained that $N_\mathrm{e} \approx 29$, which corresponds to the yellow region in Fig.~\ref{fig:thw}. Thus, it is expected that selecting several combination of $th$ and $w$ values (which correspond to the yellow region in Fig.~\ref{fig:thw}) can lead to correct estimation of the number of ejected baby droplets. In the present study, $th=19$\% and $w=17$ pixel were selected. Similar results are obtained for the rest of the tested conditions.

\section*{Appendix C: Selection of time required for estimation of $K_1$}

In section~\ref{atomcont}, it is argued that $D^2/D^2_0$ data corresponding to 90\% of the droplet lifetime (i.e. $\Delta t = 0.9\tau$) was used for estimation of $K_1$. For this estimation, the time instant corresponding to the median of the data was set to $t = 0.5\tau$. For all tested conditions, $\Delta t = 0.9\tau$ was used to avoid nonlinear variation of $D^2/D^2_0$ corresponding to the droplet heat-up and flame extinction periods. A smaller number of data points (corresponding to smaller value of $\Delta t/\tau$) can also be used for estimation of $K_1$. However, as it is argued in the following, $\Delta t = 0.9\tau$ leads to the most accurate estimation of $K_1$.

The quality of the linear fits was used to assess the value of $\Delta t$ required for the calculation of $K_1$. Two criteria were used to assess this quality, similar to \cite{bennewitz2019systematic}. Specifically, the coefficient of determination ($R^2$) and the linear regression coefficient ($\beta$) were estimated using:

\begin{subequations}
	\begin{equation}
		R^2(\Delta t)= 1 - \dfrac{\sum_{i=1}^{N(\Delta t)}[D^2_i-\overline{D^2}(\Delta t)]^2}{\sum_{i=1}^{N(\Delta t)}(D^2_i-f_i)^2},
		\label{eq:R2}
	\end{equation}
	\begin{equation}
		\beta = R^2 \frac{\Delta t}{\tau},
		\label{eq:beta}
	\end{equation}
\end{subequations}
where $i$ is the number of the utilized data points, $N$ is the total number of data points corresponding to $\Delta t$, $D^2_i$ is the experimentally measured data (i.e., the equivalent droplet diameter squared), $\overline{D^2}$ is the mean of $D^2_i$, and $f_i$ is the value of $D^2$ obtained from a linear fit. For a given $\Delta t$, this fit was obtained using the least square technique. Both values of $R^2$ and $\beta$ being close to unity suggest that the selected $\Delta t$ allows for accurate estimation of $K_1$, and that the fit has an acceptable quality.

\begin{figure}
	\centering
	\includegraphics[width=0.5\textwidth]{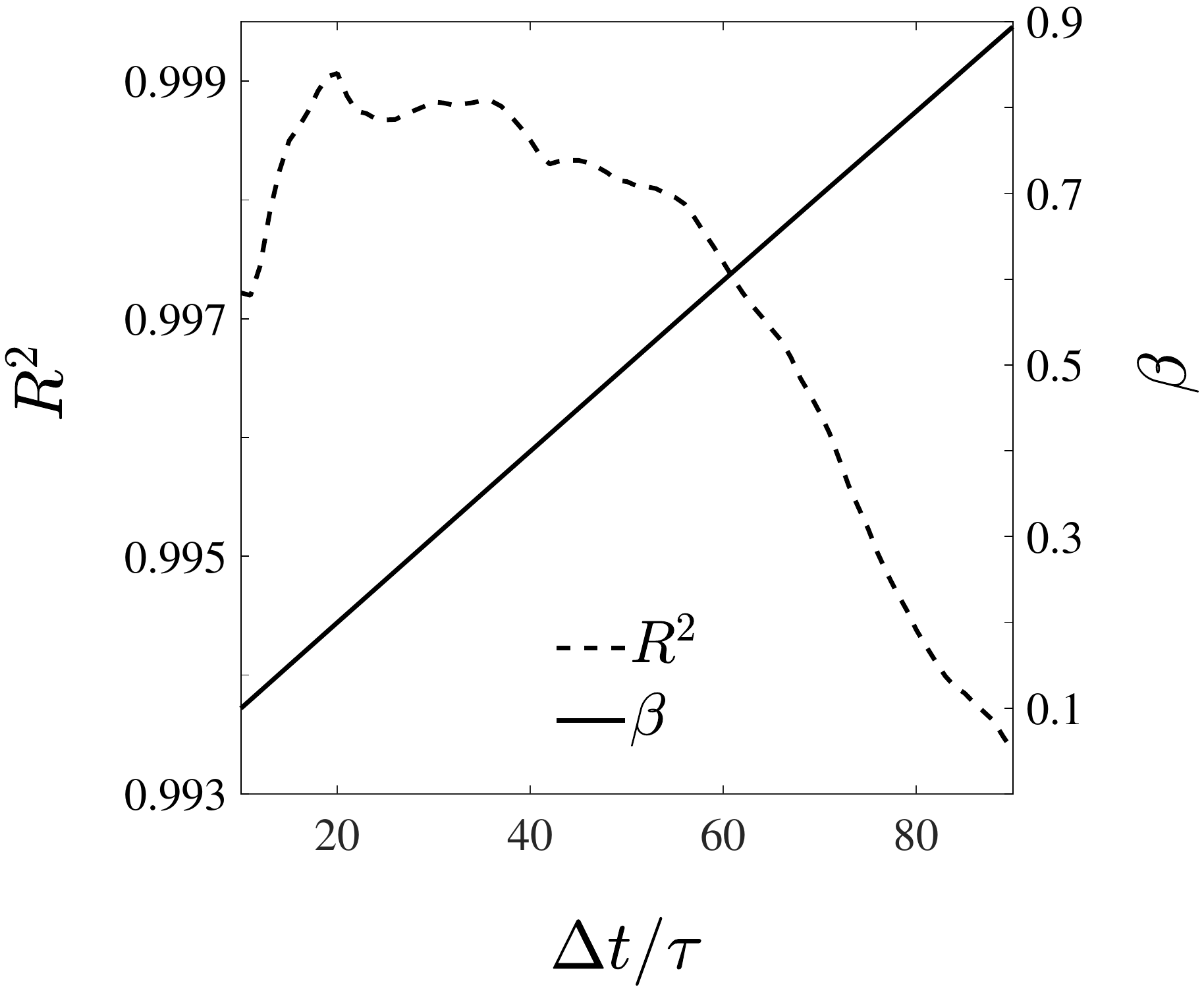} 
	\caption{Variations of $R^2$ and $\beta$ obtained from Eqs.(\ref{eq:R2})~and~(\ref{eq:beta}) versus $\Delta t/\tau$. The results are presented for the representative tested condition of Ep0.01.}
	\label{fig:90}
\end{figure}

Values of $R^2$ and $\beta$ were obtained for several $\Delta t/\tau$ using Eqs.~(\ref{eq:R2})~and~(\ref{eq:beta}), and the results are presented in Fig.~\ref{fig:90} using dashed and solid black curves, respectively. The results correspond to Ep0.01 test condition. Similar results were obtained for the rest of the tested conditions. As can be seen, the values of $R^2$ are relatively large and close to unity; however, $\beta$ maximizes for $\Delta t= 0.9 \tau$. Thus, 90\% of $D^2/D^2_0$ data were used for calculation of $K_1$.

\bibliographystyle{abbrv}
\bibliography{FullPaper}

\end{document}